\documentclass[fleqn,usenatbib]{mnras}
\pdfoutput=1 
\usepackage{newtxtext,newtxmath}

\usepackage[T1]{fontenc}

\DeclareRobustCommand{\VAN}[3]{#2}
\let\VANthebibliography\thebibliography
\def\thebibliography{\DeclareRobustCommand{\VAN}[3]{##3}\VANthebibliography}

\usepackage{graphicx}	
\usepackage{amsmath}	
\usepackage{enumerate}
\usepackage{multirow}
\usepackage{longtable}
\usepackage{color, colortbl}
\usepackage[table]{xcolor} 
\usepackage{caption}
\usepackage{subcaption}

\title[planet formation in perturbed discs]{The role of density perturbation on planet formation by pebble accretion}

\author[Andama et al.]{
G. Andama,$^{1}$\thanks{E-mail: gandama@must.ac.ug}
N. Ndugu,$^{2,3}$\thanks{E-mail: n.ndugu@muni.ac.ug}
S.K. Anguma,$^{3}$
E. Jurua$^{1}$\thanks{E-mail: ejurua@must.ac.ug}
\\
$^{1}$Department of Physics, Mbarara University of Science and Technology, Mbarara, Uganda\\
$^{2}$Department of Physics, North-West University, Private Bag X2046, Mmabatho 2735, South Africa\\
$^{3}$Department of Physics, Muni University, Arua, Uganda\\
}

\date{Accepted 2022 March 15. Received 2022 March 13; in original form 2022 January 28}

\pubyear{2022}

\begin{document}
\label{firstpage}
\pagerange{\pageref{firstpage}--\pageref{lastpage}}
\maketitle
\numberwithin{equation}{section}
\numberwithin{figure}{section}
\begin{abstract}
Protoplanetary discs exhibit a diversity of gaps and rings of dust material, believed to be a manifestation of pressure maxima commonly associated with an ongoing planet formation and several other physical processes. Hydrodynamic disc simulations further suggest that multiple dust ring-like structures may be ubiquitous in discs. In the recent past, it has been shown that dust rings may provide a suitable avenue for planet formation. We study how a globally perturbed disc affects dust evolution and core growth by pebble accretion. We performed global disc simulations featuring a Gaussian pressure profile, in tandem with global perturbations of the gas density, mimicking wave-like structures, and simulated planetary core formation at pressure minima and maxima. With Gaussian pressure profiles, grains in the inside disc regions were extremely depleted in the first 0.1 Myrs of disc lifetime. The global pressure bumps confined dust material for several million years, depending on the strength of perturbations. A variety of cores formed in bumpy discs, with massive cores at locations where core growth was not feasible in a smooth disc, and small cores at locations where massive cores could form in a smooth disc. We conclude that pressure bumps generated by a planet and/or other physical phenomena can completely thwart planet formation from the inside parts of the disc. While inner disc parts are most favourable for pebble accretion in a smooth disc, multiple wave-like pressure bumps can promote rapid planet formation by pebble accretion in broad areas of the disc.

\end{abstract}




\section{Introduction}\label{sec:introduction}

A plethora of physical processes takes place in protoplanetary discs that sets the stage for the formation of planetary bodies~\citep[for a comprehensive review, see][]{armitage2019}. 
The two main approaches that have been used extensively in studying the physical process of planet formation are the top-down and bottom-up theories. 

The top-down theory is concerned with planet formation via the gravitational collapse of sufficiently massive gaseous disc~\citep{kuiper1951, cameron1978, boss1997,gammie2001,rice2003,tanga2004,rafikov2005,durisen2006}. One major limitation of this approach is that it mainly produces giant planets, although tidal downsizing may follow, resulting in much smaller
planets ~\citep{nayakshin2010}. 
Another limitation of the gravitational collapse is that it occurs preferably in the outer disc locations where the gas is dynamically cold enough to allow this process~\citep[][]{boss1997,boley2009,armitage2010,humphries2019}. Thus it is difficult to explain the existence of the less massive planets in the inner disc parts based on the theory of gravitational collapse.

A more natural way to explain the existence of the Solar system and several other planetary systems is the bottom-up model~\citep{safronov1969}, in which growth starts from the smallest dust grains, all the way to the larger kilometre-sized bodies called planetesimals. Planetary cores are then formed when the larger planetesimals $\ge 100$ km in size gravitationally attract the smaller ones through a mechanism commonly referred to as core accretion of planetesimals~\citep[][]{wetherill1980,kokubo1998,thommes2003,coleman2014}. The formation of planetary cores by core accretion of planetesimals is typically slow~\citep[][]{tanaka1999,thommes2003,levison2010,johansen2019b}, but growth from less than 1 km-size planetesimals may still be fast~\citep[e.g.,][]{mordasini2009}. However, until now, there are no pieces of evidence in the Solar system for planetesimals of such smaller sizes~\citep[][]{bottke2005,bottke2005b, morbidelli2009,singer2019}. In addition, planetary core formation by accretion of planetesimals alone cannot satisfactorily explain many aspects of the observed planetary systems. 

An alternative, more attractive core formation pathway where the core accretes small bodies typically in the millimetre-centimetre size range assisted by gas drag, referred to as pebble accretion, was born~\citep{johansen2010, ormel2010,lambrechts2012,lambrechts2014a}. In general, these small solid bodies, often called pebbles, are better designated by their aerodynamic property called the Stokes number, which defines the degree to which they are coupled to gas and hence affected by drag forces. Pebble accretion has become more attractive in studying the formation of planetary bodies because it can, in principle, explain broad, if not all, aspects of both the Solar and exoplanetary systems in a much more natural manner than the previous methods. However, it is becoming increasingly clear that planets can form by hybrid accretion of both pebbles and planetesimals~\citep{alibert2018,guilera2020,venturini2020b,izidoro2021}.

The pebble accretion paradigm suffers from the fact that centimetre-sized solids are lost to the central star as they rapidly drift inward on short dynamical timescales~\citep{whipple1972,weidenschilling1977,takeuchi2005, alexander2007,brauer2007, brauer2008}.  In addition, the mm-cm sized pebbles, with Stokes number greater than 0.1, may even be lost on similar timescales~\citep{johansen2019}.  Consequently, this affects planet formation because, first and foremost, the mm-cm size pebbles are needed to form planetesimals via gravitational collapse, some of which form planetary embryos~\citep{johansen2007}. Secondly, the planetary embryos also need pebbles to grow into planets by pebble accretion. Thus, planetesimals may not form while the core may not afford to grow if pebbles get drained in the manner explained above.

While most discs seem to have lost considerable amounts of pebbles, per theoretical predictions, as suggested by the recent studies on disc surveys~\citep{tychoniec2020}, a substantial amount of dust grains in the mm – cm range seem to survive in a few discs that have evolved good enough~\citep{testi2003,wilner2005,rodmann2006,brauer2007,perez2012,trotta2013,carrasco2016,ansdell2017}, which is still puzzling

A few studies have fronted some classic solutions to grain loss via radial drift to explain the possibility of grain survivability in some discs.  For example, a destructive collision of larger dust aggregates may produce smaller grains~\citep{blum2008}, which may then be retained in the disc for a good period since they drift less rapidly compared with the larger grains. Another possible mechanism is the coagulation-fragmentation equilibrium~\citep{dominik2008} which can eventually lead to grain retention, as demonstrated by \cite{birnstiel2009}. There are other physical effects such as zonal flows formed by magnetorotational instability (MRI)~\citep[e.g.,][]{johansen2009,dzyurkevich2010,johansen2011,uribe2011}, which may induce pressure bumps across the gas disc and  therefore trap and retain dust grains in the disc~\citep{pinilla2012}.

From the ALMA data, discs exhibit a diversity of gaps and dust ring structures~\citep{huang2018,long2018,vandermarel2019}, which provide evidence for pressure bumps. Although some studies attribute the gaps and dust rings to gravitational interactions of forming planets with their natal discs~\citep[e.g.,][]{wolf2005,dodson2011,zhu2011,gonzalez2012,ataiee2013,perez2015,dipierro2015,dong2015,zhu2015,fung2015,picogna2015,bae2016,kanagawa2016,rosotti2016,dong2017,isella2018,pinte2018,teague2018,dong2018, zhang2018,dullemond2018b}, it is still highly debatable as to whether all these observed gaps and dust rings are signposts of planet formation~\citep{lodato2019, ndugu2019, nayakshin2019, morbidelli2020}.

Other proposed possible causes of the ring structures include: the inversion of density profile at snowline~\citep{banzatti2015,zhang2015,okuzumi2016,pinilla2017}; dust feedback on the gas~\citep{takahashi2014,takahashi2016,gonzalez2017,dullemond2018a,tominaga2020}; confinement of dust grains by magnetic fields in the disc~\citep{bai2014,simon2014}; transition of density structure in the deadzones~\citep{regaly2012,flock2015,flock2016,flock2017,lyra2015}. Furthermore, infall of material from star-forming environment could trigger formation of substructures in the form of rings, gaps, and spirals~\citep{kuznetsova2022}.

No matter how the pressure bumps form, it is now well understood that they cut off the inward flow of solid material, which eventually piles at the pressure maxima. Consequently, this impedes pebble accretion in regions interior to the pressure bump~\citep{izidoro2021}. In addition, the pile-up of dust grains may reach overdensities at pressure maxima locations, with the possibility of facilitating planetesimal formation~\citep{eriksson2020,shibaike2020}, which may turn out to be potential targets for accretion by a nearby planet~\citep{guilera2020, izidoro2021}. However, large orbital excitations may expel planetesimals from their birth locations before they are accreted~\citep{eriksson2021}. Nevertheless, some dust grains may undergo turbulent diffusion through the pressure bump, depending on the turbulence strength and pressure maxima~\citep{zhu2012,pinilla2016,weber2018, bitsch2018,haugbolle2019}.

The presence of pressure bumps in the disc has led to a paradigm shift in the pebble accretion scenario, where the formation
of planets within such structures has become particularly interesting~\citep{morbidelli2020,guilera2020,izidoro2021}. For
example,~\cite{morbidelli2020} studied the formation of a planet by assuming that the growing planetary embryo is locked in the pressure bump and only accretes material from the dust ring in which it is embedded. In this scenario, instead of the classical pebble isolation mass~\citep{lambrechts2014a}, the authors derived the final planetary core mass from the total mass of pebbles in the dust rings. Furthermore, the authors also found that the formation of planetary cores is rapid inside the pressure bump and that the final core mass may be comparatively smaller than that fixed by the pebble isolation mass.

~\cite{guilera2020} simulated how a gas giant planet would form at a pressure bump induced by viscosity transition at the water-ice line. The planetary core grows, first by accretion of pebbles until pebble isolation mass, followed by accretion of planetesimals formed from the pebble pile-up at the pressure bump. As the case for the dust rings,~\cite{guilera2020} reported fast planet formation at the water-ice line in the presence of a pressure maximum. On the other hand, the work of~\cite{izidoro2021} assumes a similar pressure bump structure as in~\citet{pinilla2012},~\cite{dullemond2018b} and~\cite{morbidelli2020}, where the authors studied how a strong pressure bump changes the picture of terrestrial planet formation in the framework of the Solar system. Because the pebble flux is cut off by the pressure bump,~\cite{izidoro2021} concluded that the terrestrial planets formed from planetesimals rather than by pebble accretion.

However, the previous works discussed above neglected how global perturbations in the gas density profile, and dust evolution in the perturbed disc would affect planet formation in such an environment. We think that changes in gas density as discussed above ultimately results in plenty of dust material in some parts of the disc and dearth of material in other parts, which motivates us to address a couple of questions. Suppose long-lived zonal flows exist that lock dust material in the disc. How do they change the overall picture of core accretion? Can global dust evolution save planet formation via pebble accretion by replenishing grains interior to a gap opened by a planet and other mechanisms? 

The goal of this paper is, therefore, to numerically explore how global perturbations in the disc structure and a gap taken together affect growth and evolution of solid material and what this means for core growth by pebble accretion. Here, we use the approach of~\cite{pinilla2012} for global perturbations in the disc structure and~\cite{dullemond2018b} for dust evolution in a disc with a gap opened by a planet or other mechanisms. In particular, we simulate the evolution of dust material in the context of three simple scenarios: (a) smooth density profile with a single gap, (b) sinusoidal perturbations without a gap in the disc and (c) sinusoidal perturbations with a gap in the disc. We then study,  in a self-consistent manner, how these different scenarios can affect core growth via pebble accretion. Our simulations take into account full grain growth, fragmentation and drift limits.

The rest of the paper is structured as follows: In Section~\ref{theory}, we describe the underlying disc model, density perturbation model and the core growth model. We present and discuss our results in Section~\ref{results}. We then summarise our findings in Section~\ref{conclusions}.

\section{Method}\label{theory}
\subsection{Disc model}
In this work, we use the 1D viscous gas evolution and the two population dust evolution model of~\cite{birnstiel2012}. The gas evolves viscously as~\citep{hueso2005,birnstiel2009}
\begin{equation}
 \frac{\partial\Sigma_{\rm{g}}}{\partial t}=\frac{1}{r}\frac{\partial}{\partial r}\left( \Sigma_{\rm{g}}ru_{\rm{g}} \right),\label{eq:01}
\end{equation}
where $\Sigma_{\rm{g}}$ is the gas surface density, $r$ is the radial distance and $u_{\rm{g}}$ is the radial gas velocity given by
\begin{equation} 
 u_{\rm{g}}=-\frac{1}{\Sigma_{\rm{g}}\sqrt{r}}\frac{\partial}{\partial r} \left(\Sigma_{\rm{g}}\nu_{\rm{g}}\sqrt{r} \right).\label{eq:02}
\end{equation}
In equation~(\ref{eq:02}), $\nu_{\rm{g}}$ is the turbulent viscosity of gas which is described as~\citep{pringle1981}
\begin{equation}
 \nu_{\rm{g}}=\alpha_{\rm{t}}c_{\rm{s}}h_{\rm{g}}, \label{eq:03}
\end{equation}
where $\alpha_{\rm{t}}$ is the turbulence strength~\citep{shakura1973} and $h_{\rm{g}}$ the pressure scale height given by
\begin{equation}
 h_{\rm{g}}=\frac{c_{\rm{s}}}{\Omega_{\rm{K}}}.\label{eq:04}
\end{equation}
Here, $\Omega_{\rm{K}}$ is the Keplerian frequency given by
\begin{equation}
 \Omega_{\rm{K}}=\sqrt{\frac{GM_\star}{r^{3}}} \label{eq:05}
\end{equation}
where $G$ is the gravitational constant and $M_\star$ the mass of the central star.

The isothermal sound speed $c_{\rm{s}}$  is calculated from
\begin{equation}
 c_{\rm{s}}=\sqrt{\frac{k_{\rm{B}}T}{\mu m_{\rm{p}}}},\label{eq:06}
\end{equation}
where $k_{\rm{B}}$ is the Boltzmann constant, $T$ the mid-plane temperature, $\mu$ the mean molecular weight and $m_{\rm{p}}$ the proton mass. Assuming a Solar-mass star, we calculate the mid-plane temperature using a simple power law  for Minimum Mass Solar Nebula (MMSN) model \citep{hayashi1981} as 
\begin{equation}
 T=280\left(\frac{r}{\rm{au}} \right)^{-1/2},\label{eq:07}
\end{equation}
where we assume stellar irradiation dominates over viscous heating.

In the simulations, the initial gas surface density $\Sigma_{\rm{g,0}}$ closely follows the self-similar solution of~\cite{lyndenbell1974} as in~\cite{drazkowska2021} given by
\begin{equation}
\Sigma_{\rm{g,0}}(r)=\frac{M_{\rm{disc}}}{2\pi r_{\rm{c}}^{2}} \left( \frac{r}{r_{\rm{c}}} \right)^{-p}{\rm{exp}}\left[-\left( \frac{r}{r_{\rm{c}}} \right)^{2-p} \right]. \label{eq:08}
\end{equation}
Here, $M_{\rm{disc}}$ is the initial disc mass,  $p$ is the viscosity power-law index and $r_{\rm{c}}$ the characteristic radius at an initial time $t_{0}$.

The dust evolution in~\cite{birnstiel2012} features full grain size distribution regulated by grain growth, fragmentation and drift limits. In this study, we modify the dust evolution routine where the dust surface density $\Sigma_{\rm{d}}$ for the two grain populations evolves as~\citep[Equation~A.3 ][]{birnstiel2012}
\begin{equation}
 \frac{\partial\Sigma_{\rm{d}}}{\partial t}+\frac{1}{r}\frac{\partial}{\partial r}\left[r\left(\Sigma_{\rm{d}}{u^{*}}-D^{*}\Sigma_{\rm{g}}\frac{\partial}{\partial r}\left(\frac{\Sigma_{\rm{d}}}{\Sigma_{\rm{g}}} \right) \right) \right]=0, \label{eq:09}
\end{equation}
where
\begin{align}
u^{*} &=\bar{u}-(D_{0}-D_{1})\frac{\partial f_{\rm{m}} }{\partial r},\label{eq:10}\\
D^{*}&=(D_{0}-D_{1})f_{m}+D_{1}.\label{eq:11}
\end{align}
Here, $\bar{u}$ is the mass weighted radial drift velocity of the dust component and $f_{\rm{m}}$ the mass fraction of the two dust populations. Both $\bar{u}$ and $f_{\rm{m}}$ are calculated as in~\cite{birnstiel2012}. $D_{0}$ and $D_{1}$ are the dust diffusivities of the small and large population, respectively given by~\citep{youdin2007}
\begin{equation}
 D_{0}=\frac{\nu_{\rm{g}}}{1+\tau_{0}}~\text{and }D_{1}=\frac{\nu_{\rm{g}}}{1+\tau_{1}}. \label{eq:12}
\end{equation}
Here, $\tau_{0}$ and $\tau_{1}$ are the Stokes number of the small or large dust population, respectively as described in~\cite{birnstiel2012}.
The surface densities $\Sigma_{0}$ and $\Sigma_{1}$ for the small and large size populations respectively are then calculated from
\begin{align}
 \Sigma_{0}&=(1-f_{\rm{m}})\Sigma_{\rm{d}},\label{eq:13}\\
 \Sigma_{1}&=f_{\rm{m}}\Sigma_{\rm{d}}.\label{eq:14}
\end{align}

\subsection{Pressure bump model}
In this section, we describe the different pressure bump models used in our work. These include (a) a sinusoidal pertubation in gas density that mimics processes similar to zonal flows (b) a single pressure bump that mimics perturbation of gas surface density by a planet and (c) a disc with both the sinusoidal perturbations and a gap. We then study grain evolution in all these cases. 
\subsubsection{Sinusoidal perturbation}
We performed self-consistent dust evolution with full grain growth, fragmentation and drift limits in the perturbed disc structures.
To mimic pressure bumps induced by some physical effects, we followed~\cite{pinilla2012} and introduced a sinusoidal perturbation $F_{\rm{wave}}$ defined by
\begin{equation}
 F_{\rm{wave}}=1+A\cos\left(\frac{2\pi r}{L(r)} \right), \label{eq:15}
\end{equation}
where A and $L(r)$ are the amplitude and wavelength of the sinusoidal perturbation.
The perturbed gas surface density is obtained by modifying the unperturbed initial gas density in Equation (\ref{eq:08}) as
\begin{equation}
 \tilde\Sigma_{\rm{g,0}}=F_{\rm{wave}}\Sigma_{\rm{g,0}}\label{eq:16}. 
\end{equation}
We then calculated the perturbed initial dust surface density as
\begin{equation}
 \tilde\Sigma_{\rm{d,0}}=Z\tilde\Sigma_{\rm{g,0}},\label{eq:17}
\end{equation}
where $Z$ is the initial dust-to-gas ratio.

The stability of hydrostatic equilibrium requires that  perturbation  wavelengths be greater than the gas pressure scale heights~\citep{pinilla2012,dullemond2018b}. We therefore set 
\begin{equation}
 L(r)=f h_{\rm{g}}(r), \label{eq:18}
\end{equation}
where $f$ is the scaling factor for the wavelength which we assume to take the range $1\le f\le 3$ as in \cite{pinilla2012}.
Furthermore, to guarantee disc stability, requires that the wave amplitude lies in the range $0.1\le A \le 0.35$~\citep{pinilla2012}.
\subsubsection{Disc with gaps}
The gas surface density may also be modified locally by a growing planet when it opens a gap in the disc or by viscosity transition at water-ice line, which can occur when condensation of icy grains removes free electrons from the gas, which causes a jump in resistivity profile and hence viscosity~\citep{kretke2007,bitsch2014b,guilera2017,guilera2020}. Thus, following \cite{dullemond2018b}, we introduce a gap with width $w_{\rm{gap}}$ at some radial distance $r_{\rm{gap}}$ in the disc that is wider and deeper than troughs that arise from the global sinusoidal perturbation in Equation~(\ref{eq:15}). Thus, we set $w_{\rm{gap}}>L(r)$ such that
\begin{align}
 w_{\rm{gap}}&=f_{\rm{gap}}h_{\rm{g}}(r),\label{eq:19} \\
 f_{\rm{gap}}&=1+f,\label{eq:20}
\end{align}
where $f_{\rm{gap}}$ is a factor that defines the gap depth. We then define a Gaussian gap profile $F_{\rm{gap}}(r)$ as in~\cite{dullemond2018b}, which scales as
 \begin{equation}
  F_{\rm{gap}}(r)=\exp\left[-f_{\rm{gap}}\exp\left(-\frac{dr^{2}}{2 w_{\rm{gap}}^{2}} \right) \right],\label{eq:21}
 \end{equation}
where $dr=r-r_{\rm{gap}}$.
 We now define $\tilde\Sigma_{\rm{g,0}}^{\prime}$ as the initial gas surface density with a gap, given by
 \begin{equation}
  \tilde\Sigma_{\rm{g,0}}^{\prime}=\Sigma_{\rm{g,0}}F_{\rm{gap}}(r).\label{eq:22}
 \end{equation}
To ensure that the gaps are not smoothed out by viscosity and that they remain open during disc evolution, we use the same technique as in~\cite{dullemond2018b} where $\nu_{\rm{g}}$ is re-scaled as 
\begin{equation}
 \tilde\nu_{\rm{g}}=\frac{\nu_{\rm{g}}}{F_{\rm{gap}}(r)}.\label{eq:23}
\end{equation}

 The corresponding dust profile that is trapped by the pressure bump that accompanies the gap is given by
 \begin{equation}
  \tilde\Sigma_{\rm{d}}^{\prime}(r)=\Sigma_{\rm{d}}\exp\left(-\frac{dr^{2}}{2 w_{\rm{d}}^{2}} \right),\label{eq:24}
 \end{equation}
where dust trap width $w_{\rm{d}}$ is defined as
\begin{equation}
 w_{\rm{d}}=w_{\rm{gap}}\left(\frac{\alpha_{t}}{\tau} \right)^{1/2}, \label{eq:25}
\end{equation}
where $\tau$ is the Stokes number of the dust grains.
 For the two dust populations, we can write the corresponding perturbed surface densities $\tilde\Sigma_{0}$ and $\tilde\Sigma_{1}$ as
 \begin{align}
  \tilde\Sigma_{0}^{\prime}&=\Sigma_{0}\exp\left[-\exp\left(-\frac{dr^{2}}{2 w_{\rm{d,0}}^{2}} \right) \right],\label{eq:26}\\
  \tilde\Sigma_{1}^{\prime}&=\Sigma_{1}\exp\left[-\exp\left(-\frac{dr^{2}}{2 w_{\rm{d,1}}^{2}} \right) \right].\label{eq:27}
 \end{align}
Here $w_{\rm{d,0}}$ and $w_{\rm{d,1}}$ are the respective dust ring widths for each population which we write as
\begin{align}
 w_{\rm{d,0}}=w_{\rm{gap}}\left(\frac{\alpha_{t}}{\tau_{0}} \right)^{1/2} \rm{ and~}
 w_{\rm{d,1}}=w_{\rm{gap}}\left(\frac{\alpha_{t}}{\tau_{1}} \right)^{1/2}. \label{eq:28}
\end{align}
Therefore, the initial dust surface density with gap in Equation~(\ref{eq:24}) modifies the two dust populations to
\begin{equation}
 \tilde\Sigma_{\rm{d}}^{\prime}(r)=\tilde\Sigma_{0}^{\prime}+\tilde\Sigma_{1}^{\prime}.\label{eq:29}
\end{equation}

 \subsubsection{Disc with both sinusoidal perturbation and gaps}
 We now reconfigure the disc such that it has both a sinusoidal gas density profile and a gap by combining Equations~(\ref{eq:15}) and~(\ref{eq:21}) as
 \begin{equation}
  F(r)=F_{\rm{wave}}(r)F_{\rm{gap}}(r). \label{eq:30}
 \end{equation}
The modified gas density profile $\tilde\Sigma_{\rm{g,0}}^{\prime\prime}$ is then given by
\begin{equation}
 \tilde\Sigma_{\rm{g,0}}^{\prime\prime}=F(r)\Sigma_{\rm{g,0}}
\end{equation}
and the turbulent viscosity is re-scaled as
\begin{equation}
 \tilde\nu_{\rm{g}}^{\prime\prime}=\frac{\nu_{\rm{g}}}{F(r)}.
\end{equation}

Similarly, for the initial dust surface density, we re-write Equation~(\ref{eq:29}) as
\begin{equation}
 \tilde\Sigma_{\rm{d}}^{\prime\prime}(r)=F_{\rm{wave}}(r)\tilde\Sigma_{\rm{d}}^{\prime}(r).
\end{equation}

\subsection{Planet formation model}
In this study we limit ourselves to planet formation at core stage, where the planetary core grows by pebble accretion. Pebble accretion occurs when pebbles within the gravitational influence of the core undergo gas drag and sediment to the core~\citep{ormel2010,lambrechts2012}.
 
 A protoplanet may accrete pebbles in two growth regimes, namely, the Bondi and Hill regimes~\citep{lambrechts2012}. In the Bondi accretion, also known as the drift limited accretion regime, the gravitational influence of the host star and coriolis effects are not accounted for and growth in this regime is usually slow. In the Hill regime the capture cross-section is set by the Hill radius and the protoplanet accretes pebbles more efficiently~\citep{lambrechts2012,lambrechts2014b}. 

 In our simulations, we start pebble accretion at the transition mass, which is the mass at which accretion transitions from Bondi to Hill regime~\citep{lambrechts2012}. We follow accretion model described in~\cite{andama2021} where the planetary core grows by concurrent accretion of different pebble species. The contribution of each pebble species to core growth is calculated via the classical pebble accretion formulation as~\citep{morbidelli2015}
 \begin{equation}
 \dot{M}_{\rm{2D}}=\left \{
 \begin{array}{ll}
  2\left(\tau_{\rm{i}}/{0.1} \right)^{2/3}\Omega_{\rm{K}} r_{H}^{2}\Sigma_{\rm{p,i}}& (\tau_{\rm{i}} < 0.1)\\ \\
  2\Omega_{\rm{K}} r_{H}^{2}\Sigma_{\rm{p,i}}&(\tau_{\rm{i}} \ge 0.1)
 \end{array}
 \right. \label{eq:06}
\end{equation}
where  $r_{\rm{H}}$ is the Hill radius. $\Sigma_{\rm{p,i}}$ and $\tau_{\rm{i}}$ are respectively the surface density and Stokes number of the $i$-th pebble species. $\dot{M}_{\rm{2D}}$ is the accretion in 2-D regime when scale height $h_{\rm{i}}$ of the $i$-th pebble species is less than the effective accretion radius of the planet.
 
 The planet accretes in 3-D when the pebble scale height is greater than the effective accretion radius and swicthes from 3-D to 2-D accretion as in~\cite{morbidelli2015} when
 \begin{equation}
 \dot{M}_{\rm{3D}}=\left[\sqrt{\frac{\pi}{8}}\left(\frac{\tau_{\rm{i}}}{0.1}\right)^{1/3}\frac{r_{\rm{H}}}{h_{\rm{i}}}\right]\dot{M}_{\rm{2D}}. \label{eq:07}
\end{equation}
The core accretion rate of the $i$-th pebble species is then given by 
\begin{equation}
 \dot{M}_{\rm{core,i}}=\left \{
 \begin{array}{ll}
  \dot{M}_{\rm{2D}} & \rm{for~}~\sqrt{\frac{\pi}{8}}\left(\frac{\tau_{\rm{i}}}{0.1} \right)^{1/3}r_{\rm{H}}>h_{\rm{i}}\\
  \dot{M}_{\rm{3D}}&~\text{otherwise}
 \end{array}
 \right. \label{eq:CoreGrowthRegime}
\end{equation}

If the planet grows massive enough, it may open a gap and induce a pressure bump at the outer edge of the gap, which blocks the radial drift of pebbles~\citep{morbidelli2012b}, thereby stopping pebble accretion~\citep{lambrechts2014a, bitsch2018, ataiee2018}. This mass at which the planet induces a positive pressure gradient and cuts off pebble accretion is popularly called the pebble isolation mass. The classical pebble isolation mass originally derived in~\cite{lambrechts2014a} was recently reformulated by \cite{bitsch2018} and \citet{ataiee2018}, where the isolation mass depends on both turbulent diffusion and pebble Stokes number. 

In this study we adopt the formulation in \cite{bitsch2018} as used in our previous works~\citep{andama2021,ndugu2022}. We remark here that the results of the formulations in \cite{bitsch2018} and \citet{ataiee2018} were in close agreement and we therefore think that our choice of the former formula should not qualitatively influence our results. Following~\cite{bitsch2018}, we can calculate the pebble isolation mass with turbulent diffusion, $M_{\rm{iso,i}}$, for each pebble species as
\begin{multline}
 M_{\rm{iso,i}}=17.51\left( \frac{H/r}{0.05} \right)^{3}\left[0.34\left( \frac{\rm{log\alpha_{3}}}{\rm{log\alpha_{\rm{t}}}} \right)^{4}+0.66 \right]\\
 \times\left(3.5-\frac{\partial{\rm{ln}}P}{\partial{\rm{ln}}r} \right)\left( 0.238+\frac{\alpha_{\rm{t}}}{\tau_{\rm{i}}}\right)~M_{\rm{E}}. \label{eq:PIM}
\end{multline}

The implication of Equation~\ref{eq:PIM} is that core growth continues until when the planet has grown massive enough to induce a strong pressure bump that can block pebble species with the smallest Stokes number in the grain size distribution. Hence, for planets that can grow up to pebble isolation mass, the final core mass is determined by the mass of the planet needed to block pebble species with the smallest Stokes number in the distribution.
 
Gap opening and induction of a positive pressure gradient are typical examples of planet-disc interaction.  Another example of gravitational interaction between a planet and gas disc is orbital migration whose direction depends on the net torque exerted on the planet by the disc~\citep{goldreich1979,goldreich1980,tanaka2002,paardekooper2006,baruteau2008,paardekooper2008a,paardekooper2008b,kley2008,paardekooper2009,kley2009,paardekooper2010,ayliffe2010,ayliffe2011,bitsch2011}. In this study, we focus on core growth only and therefore we implement type-I migration which better describes the orbital evolution of low mass planets.

The total torque $\Gamma_{\rm{tot}}$ that acts on the planet is calculated using the classical formula~\citep{paardekooper2011}
\begin{equation}
 \Gamma_{\rm{tot}}=\Gamma_{\rm{L}}+\Gamma_{\rm{C}},\label{torque1}
\end{equation}
where $\Gamma_{\rm{L}}$ and $\Gamma_{\rm{C}}$ are the Lindblad and corotation torques, respectively.
The Lindblad torque is calculated from
\begin{equation}
 \frac{\gamma\Gamma_{\rm{L}}}{\Gamma_{0}}=-2.5-1.7\beta+0.1s,
\end{equation}
where $\gamma=1.4$ is the adiabatic index,  $\beta$ and $s$ are the negatives of the radial gradients of temperature and gas surface density, respectively. $\Gamma_{0}$ is torque normalisation factor given by
\begin{equation}
 \Gamma_{0}=\left(\frac{q}{H/r}\right)^{2}\Sigma_{\rm{g}}r^{4}\Omega_{\rm{K}}^{2},
\end{equation}
where $q$ is the planet-star mass ratio and all quantities are calculated at planet's location.
The corotation torque originates from material corotating with the planetary body and is given by~\citet{paardekooper2010}
\begin{equation}
 \frac{\gamma\Gamma_{\rm{C}}}{\Gamma_{0}}=1.1\left(\frac{3}{2}-s \right)+7.9\frac{\xi}{\gamma}
\end{equation}
where the first and second terms are the barotropic and entropy related parts of the corotation torque. Here, $\xi=\beta-(\gamma - 1)s$ is the radial entropy gradient.

\subsection{Numerical setup}
Our numerical code incorporates the two-population code of~\cite{birnstiel2012}\footnote{\url{https://github.com/birnstiel/two-pop-py}} where we reconstructed grain size distribution using the grain size reconstruction code of~\cite{birnstiel2015}\footnote{\url{https://github.com/birnstiel/Birnstiel2015_scripts}}. 
As in~\cite{andama2021}, we ran our numerical simulations in axisymmetric 1D disc, where the computational grid extends from $r_{0}=0.05$ au to $r_{1}=3000$ au. We used a large disc with characteristic radius, $r_{\rm{c}}=200$ au and disc mass $M_{\rm{disc}}$ = 0.1 $M_{*}$. This gives a total dust mass of $\sim 330~M_{\rm{E}}$ with a nominal solid-to-gas ratio of 0.01. This dust mass is within the range of dust masses measured in different star-forming regions~\citep[see, e.g.,][]{manara2019, tychoniec2020}.

We assume a turbulent disc with $\alpha_{t}=10^{-3}$ and fragmentation velocity of grains is assumed to be 10 m/s in line with fragmentation velocities of water-ice aggregates in laboratory experiments~\citep{brauer2008,wada2008,gundlach2011,gundlach2015}. The mass of the central star was set to $M_{*}$ = 1.0 $M_{\odot}$, with temperature, $T_{*}$ = 5778 K and radius, $R_{*}$ = 1.0 $R_{\odot}$.

\begin{figure}
 \includegraphics[width=0.48\textwidth]{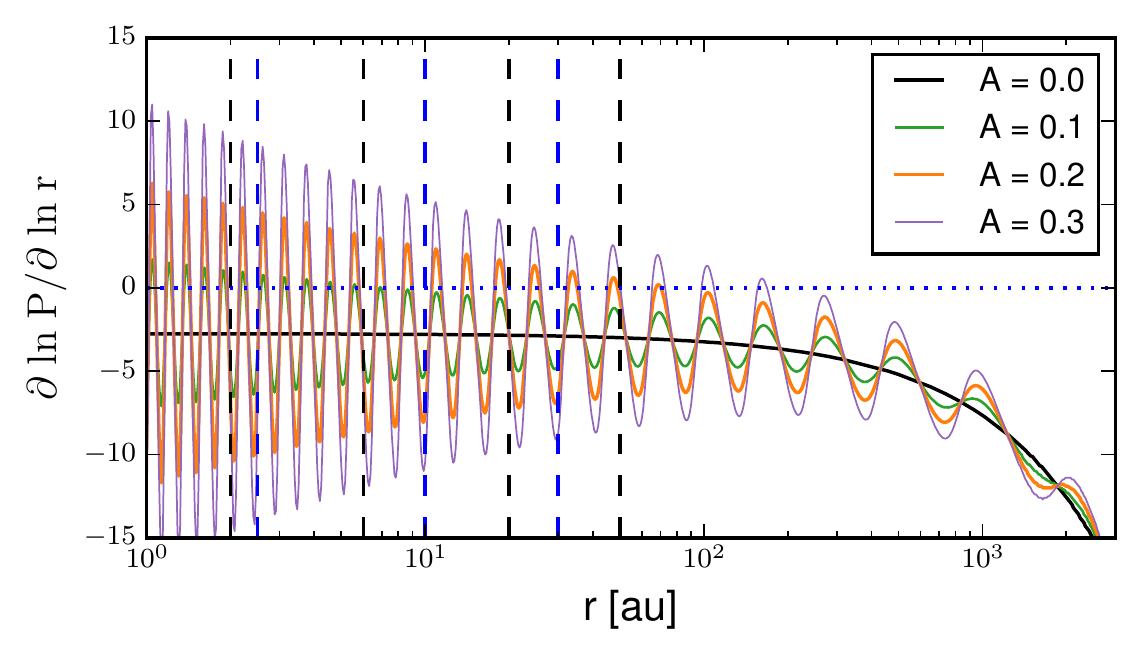}
 \caption{The pressure gradient of the perturbed disc for different perturbation amplitudes A = 0, 0.1, 0.2, 0.3 and $f$=1, where A = 0 represents unperturbed disc. Planetary embroys were implanted roughly at the pressure minima (blue dashed vertical lines) and pressure maxima (black dashed vertical lines).}
 \label{fig:fig001}
\end{figure}

We introduced perturbations in the gas density structure with width $f=1$ and amplitudes A = 0, 0.1, 0.2, 0.3. In the case of A = 0, the evolution follows that of unperturbed disc. These values of $f$ and A were selected in accordance with disc stability requirements~\citep{pinilla2012}. With this choice of parameters, the pressure profile that shapes the radial drift of dust material in our disc model is illustrated in Figure~\ref{fig:fig001}. Here, the perturbation amplitudes are higher and the widths narrower in the inner disc regions because the wavelength $L(r)$ scales with the disc scale height (see equation~(\ref{eq:18})).

We further ran simulations where we introduced local gaps bigger than the global pertubation depths at 5 au, 10 au, 50 au. It is possible that such deep gaps may be caused by early formation of planets that have reached gas accretion phase already or by other physical processes as discussed in Section~\ref{sec:introduction}. The actual location of multiple gaps requires complex modelling, which, for example, includes N-body simulations or chemical composition of the disc with ice-lines locations. These should then be anchored self-consistently to disc evolution. However, these numerical aspects are beyond the scope of this work.

We tested two scenarios, where in the first case, we performed simulations with only sinusoidal pertubations. In the second run, we include the gaps into the discs with sinusoidal perturbation and monitored dust evolution in both cases. Introduction of gaps into our disc evolution model helps us to investigate what fraction of pebbles are blocked as well as the fraction of pebbles that are able to pass through the gaps to the inner disc regions. 

Lastly, we carried out core growth in the perturbed discs where we implanted protoplanets of 0.01 $M_{\rm{E}}$ at 1 au, 2.5 au, 10 au and 30 au which lie roughly in the pressure minima, and at 2 au, 6 au, 20 au and 50 au where pressure bumps are located (see Figure~\ref{fig:fig001}). The final core masses are set by either the pebble isolation mass or the amount of pebbles trapped in the pressure bumps. The final core mass is set by pebble isolation mass if accretion is outside the pressure bump. For cores that accrete inside the pressure bump, we limit the core mass by the available mass of pebbles trapped in the pressure bump as in \cite{morbidelli2020}.

\section{Results and discussion}\label{results}

\subsection{Dust mass evolution in a perturbed disc without a gap  }\label{dust_waves}
\begin{figure*}
 \includegraphics[width=\textwidth]{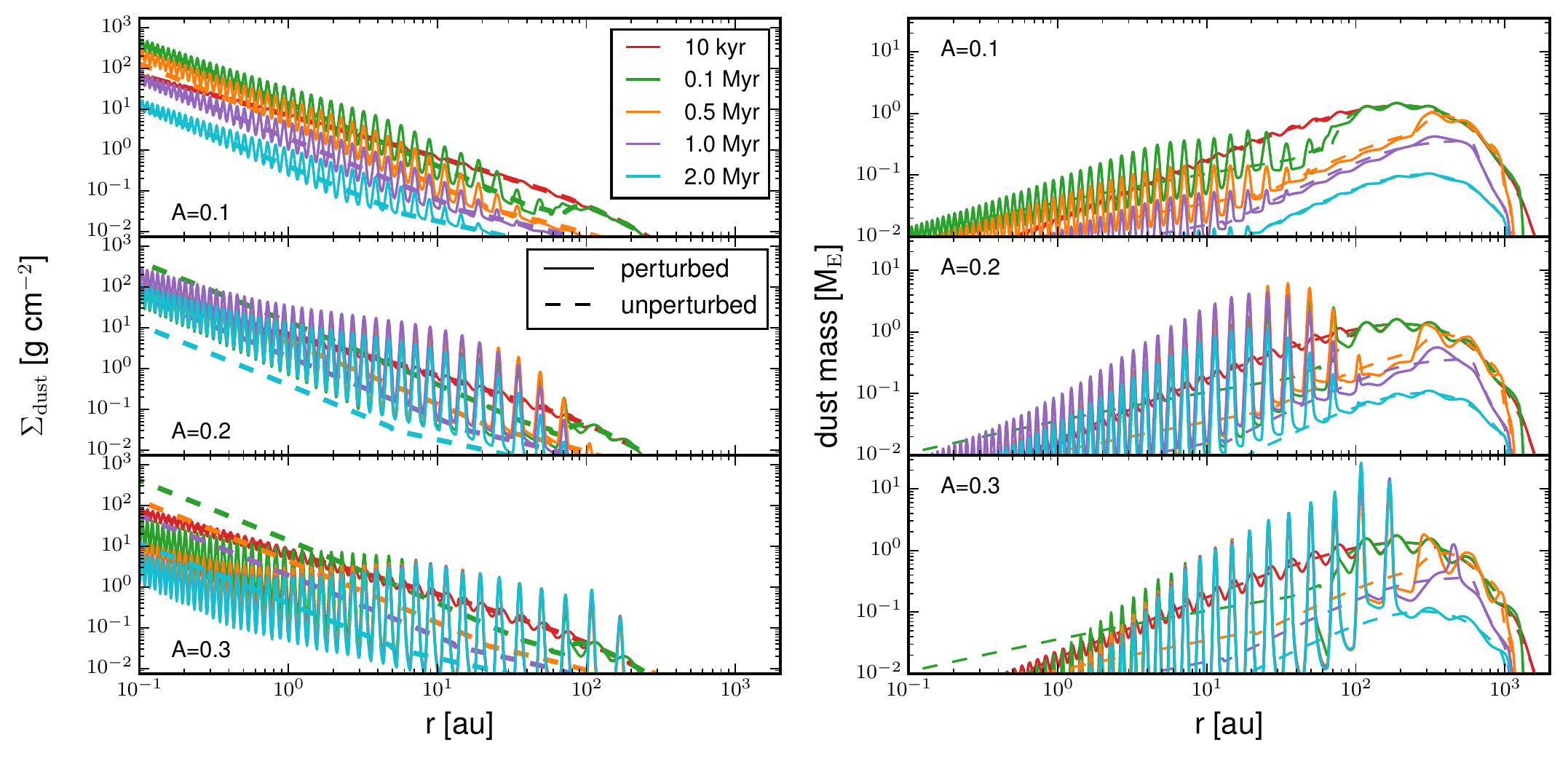}
 \caption{The evolution of dust in a disc with sinusoidal perturbation for nominal dust-to-gas ratio $f_{\rm{DG}}=0.01$, fragmentation velocity $u_{\rm{f}}=10~\rm{m/s}$ and turbulence strength $\alpha_{t}=10^{-3}$. The left panel shows the radial evolution of dust surface density for different times 10 kyr, 0.1 Myr, 0.5, Myrs, 1 Myrs and 2 Myrs. The right panel shows the corresponding radial distribution of the dust mass converted into $M_{\rm{E}}$. The dashed lines show the evolution of dust in a smooth disc.}
 \label{fig:fig002}
\end{figure*}
\begin{figure}
 \includegraphics[width=0.45\textwidth]{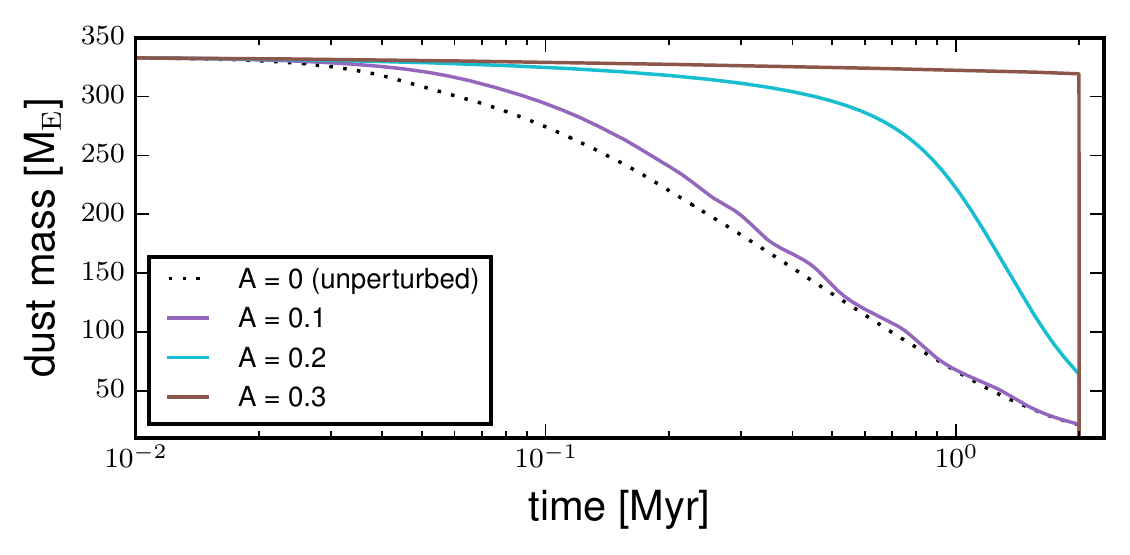}
 \caption{The time evolution of total dust mass for perturbations with wave amplitudes A = 0, 0.1, 0.2, 0.3. The total dust mass evolution for A = 0 and A =0.1 are similar, while mass decay in A = 0.2 starts after about 0.8 Myrs. For A = 0.3, there is hardly mass movement as most of the dust remains locked up in the pressure bumps.}
 \label{fig:fig003}
\end{figure}
\begin{figure*}
 \includegraphics[width=\textwidth]{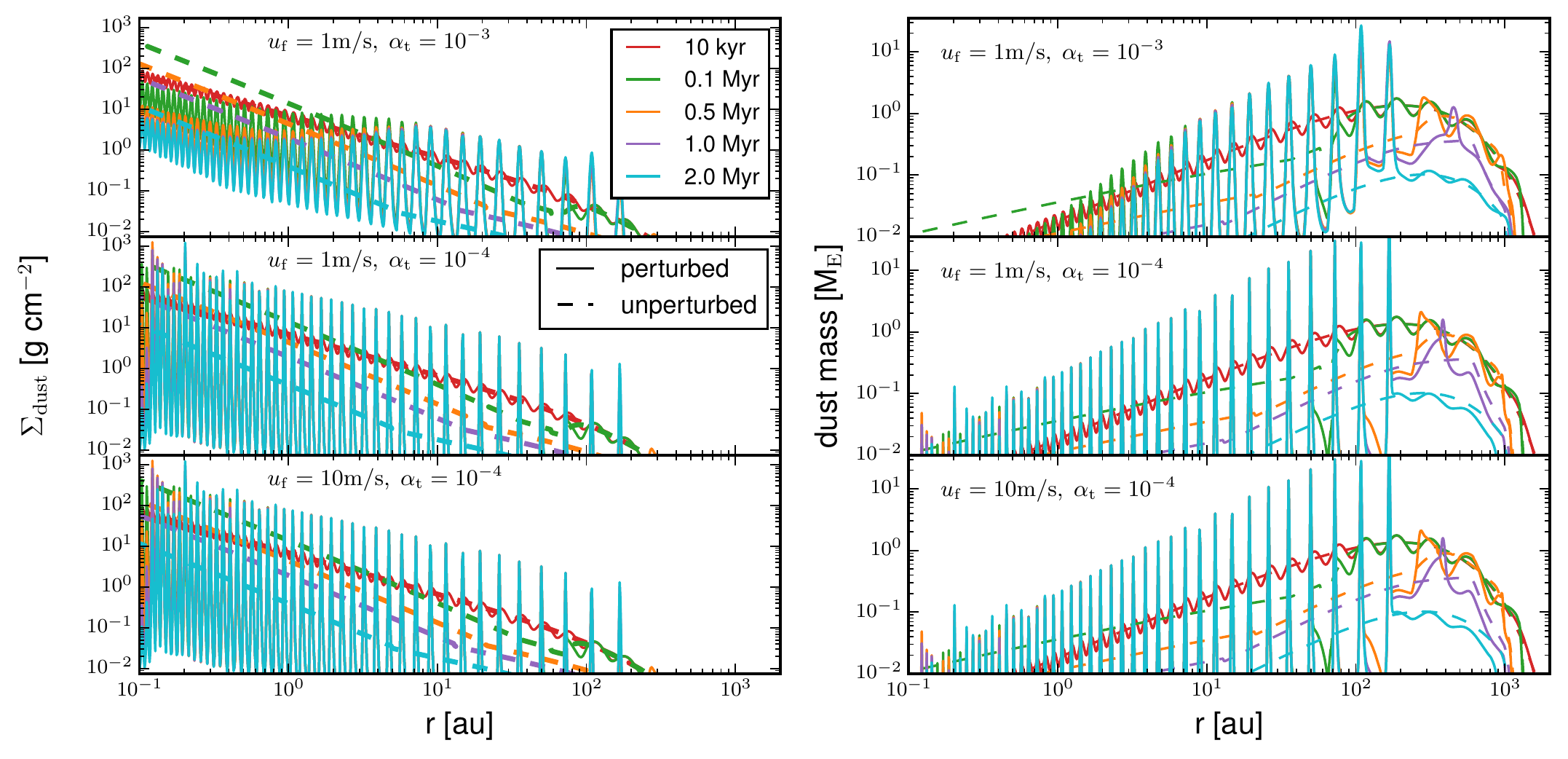}
 \caption{The evolution of dust in a strongly perturbed disc with A = 0.3 for nominal dust-to-gas ratio $f_{\rm{DG}}=0.01$ and different combinations of fragmentation velocity $u_{\rm{f}}=10~\rm{m/s}$ and turbulence strength $\alpha_{t}=10^{-3}$. The left panel shows the radial evolution of dust surface density for different times 10 kyr, 0.1 Myr, 0.5, Myrs, 1 Myrs and 2 Myrs. The right panel shows the corresponding radial distribution of the total disc mass in $M_{\rm{E}}$. The dashed lines show the evolution of dust in a smooth disc.}
 \label{fig:fig004}
\end{figure*}

In Figure~\ref{fig:fig002}, we present our results of dust evolution in  perturbed disc models where the gas density structure is modulated by a sinusoidal perturbation with different amplitudes A = 0, 0.1, 0.2, 0.3, where A = 0 represents the unperturbed profile.  The left and right panels show the evolution of the dust surface density and the radial distribution of dust mass converted into $M_{\rm{E}}$, respectively. The results of our dust evolution in the perturbed disc are qualitatively similar to that of~\cite{pinilla2012}. We discuss the main aspects of our results in the paragraphs that follow.

From the left panel of Figure~\ref{fig:fig002}, at the very early stages of disc evolution, dust evolution is very similar to that in unperturbed disc possibly because at this stage grains are just starting to drift and gather at the pressure bumps. By 0.1 Myrs, a substantial amount of dust collects at the pressure bumps, whose profile deviates from the unperturbed case because the pressure bumps continue to block more inward drifting dust material. Here, the dust evolution profile depends very much on the strength of disc perturbations.

For example, for the case of A = 0.1 as shown in the top left plot of Figure~\ref{fig:fig002}, the dust evolves in a similar fashion as the unperturbed case at each point in time. This can be connected with the dependence of grain diffusion through the pressure bumps on the grain size, the strength of turbulence and pressure maxima~\citep{zhu2012,pinilla2016,weber2018, bitsch2018,haugbolle2019}. As shown in Figure~\ref{fig:fig001}, for the perturbation amplitude A = 0.1, the pressure gradient is negative in most parts of the disc except in regions below 10 au, where the pertubation amplitudes are higher. Consequently, the dust grains experience radial drift as they easily diffuse across the weak pressure bumps~\citep{pinilla2012,pinilla2016} and hence evolve with a profile similar to that of unperturbed disc. However, the total mass trapped at the pressure bumps is more than for the corresponding radial locations in the unperturbed disc as shown in the top right panel of Figure~\ref{fig:fig002}. Furthermore, at wider orbits, for example beyond 20 au, the radial mass distribution in both unperturbed and perturbed disc are similar because the wave amplitudes are very low. 

In the middle plot on the left panel of Figure~\ref{fig:fig002}, the amplitude is set to a higher value of A = 0.2 which generates large pressure gradient variations, greatly resulting into retainment of grains at the pressure bumps. In this case, the pressure bumps lock grains where most of them are already prevented from drifting inward just after 0.1 Myr. As a result, the total amount of dust at each radial location and  at each pressure bump is then set by the amount of dust trapped at those locations. Maximal amount of dust is trapped at the pressure bumps as early as 0.1 Myr of disc evolution. For example, in the disc regions inside 10 au less than 1 $M_{\rm{E}}$ is trapped at each pressure bump while most of the dust mass is trapped between 10 -- 50 au. Here, very few earth masses of dust are locked in-between adjacent bumps as shown on the middle right panel in Figure~\ref{fig:fig002}. 

With a stronger perturbation, where A = 0.3 as shown in the bottom panel of Figure~\ref{fig:fig002}, grains are trapped at much wider orbits up to 200 au. This is because for A = 0.3, the wave amplitudes decay less slowly at wider orbital
distances compared with A = 0.1 and A = 0.2.  In the setup with A = 0.3, most of the dust mass is then locked in large parts of the disc that extends from 10 -- 200 au. In this radial domain, each pressure bump contains somewhere between 1 -- 30 $M_{\rm{E}}$ of dust, with most of the mass trapped at wider orbits. The radial drift of dust is virtually cut off for the case of A = 0.3, where dust material remains trapped within the strong pressure bumps in the entire 2 Myr of disc evolution (see Figure~\ref{fig:fig003}).

In Figure~\ref{fig:fig003}, we show how the different perturbation strengths affect the time evolution of the total disc mass. For the case of A = 0.1, the evolution of the total disc mass follows closely the profile of the unperturbed disc because dust can easily diffuse across weak pressure bumps of A = 0.1 as previously discussed.  When the perturbation amplitude is increased from A = 0.1 to A = 0.2, the total disc mass drops very slowly until 0.8 Myr after which it falls more rapidly just below 100 $M_{\rm{E}}$ at 2 Myr and most of the remaining mass is distributed between 1 -- 50 au as shown Figure~\ref{fig:fig002}. For A = 0.3, dust is trapped for the entire 2 Myrs of disc evolution, meaning there is virtually no radial drift of dust material in the disc.

For the cases of A = 0.2 and A = 0.3, dust is trapped for longer time, because initially grain growth takes place within the pressure bumps which lock up the larger grains. Since dust grains can only break through the pressure bumps if they are sufficiently small enough, the large grains must first undergo fragmentation so that they can turbulently diffuse  across the pressure bumps and get carried inward along with the gas~\citep{pinilla2016}. This process should then remove and transport dust from one pressure bump to another, but this can only be effective if fragmentation is efficient in the bumps. Since dust mass does not decay in the first 0.8 Myr for A = 0.2 and remains contant for the case of A = 0.3, this suggests that fragmentation in the pressure bumps is very slow.
Alternatively, if the grain fragmentation is efficient in the pressure bumps, the resulting smaller grains may quickly grow back to larger sizes before they diffuse away as the cycle of growth and fragmentation repeats itself in the bumps. 

This is further illustrated in Figure~\ref{fig:fig004} for the case of A = 0.3, with different combinations of fragmentation velocity and turbulent viscosity. From Figure~\ref{fig:fig004}, dust grains are more heavily trapped for the case of $\alpha_{\rm{t}}=10^{-4}$ than for $\alpha_{\rm{t}}=10^{-3}$, as shown by the very low surface densities in the pressure minima. This is because for such low disc viscosity, larger grain are produced compared with a more turbulent disc and these large grains are the most affected by the pressure bumps than the small ones. The combination of $u_{\rm{f}}=1~\rm{m/s}$ and $\alpha_{\rm{t}}=10^{-3}$ results in small size grains that are strongly coupled to the gas and drift very slowly. Hence these small grains live longer roughly close to their original locations in the disc. In this case, the dust evolution takes place over a longer period of time, where the total dust mass would not decay much as in the case of strong perturbations. This means it is their slow drift behaviour not the pressure bumps that is responsible for the longevity of these small grains.

\subsection{Dust mass evolution in a perturbed disc with a gap}\label{sec:gaps}
\begin{figure*}
 \includegraphics[width=\textwidth]{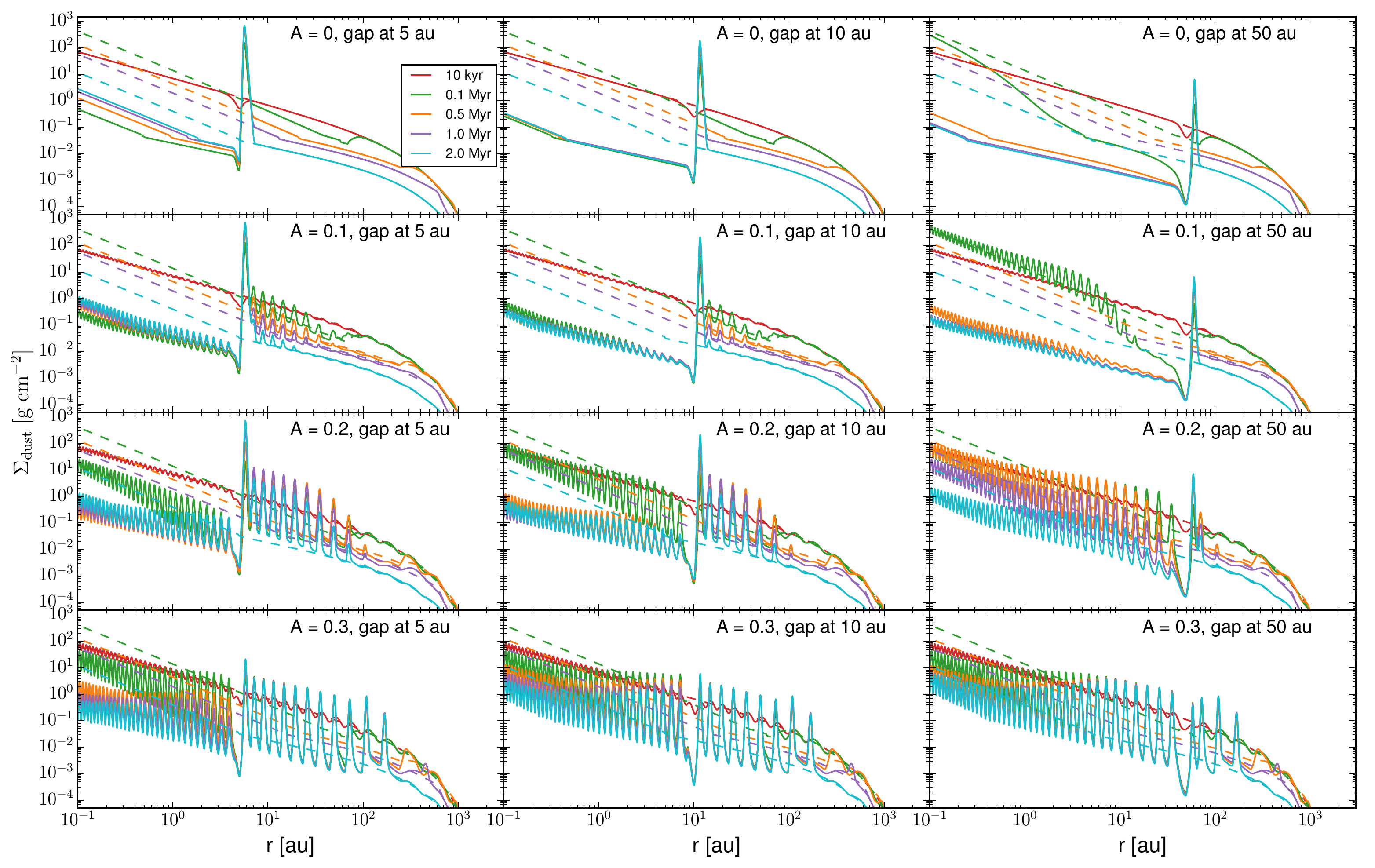}
 \caption{The evolution of dust in disc in which both a sinusoidal perturbation and a gap are introduced. The left, middle and right columns are discs with a Gaussian gap profile at 5 au, 10 au and 50 au, respectively. The first, second, third and fourth rows show discs with pertubation amplitudes A = 0 (unperturbed disc), A = 0.1, A = 0.2 and A = 0.3 respectively.}
 \label{fig:fig005}
\end{figure*}
\begin{figure}
 \includegraphics[width=0.48\textwidth]{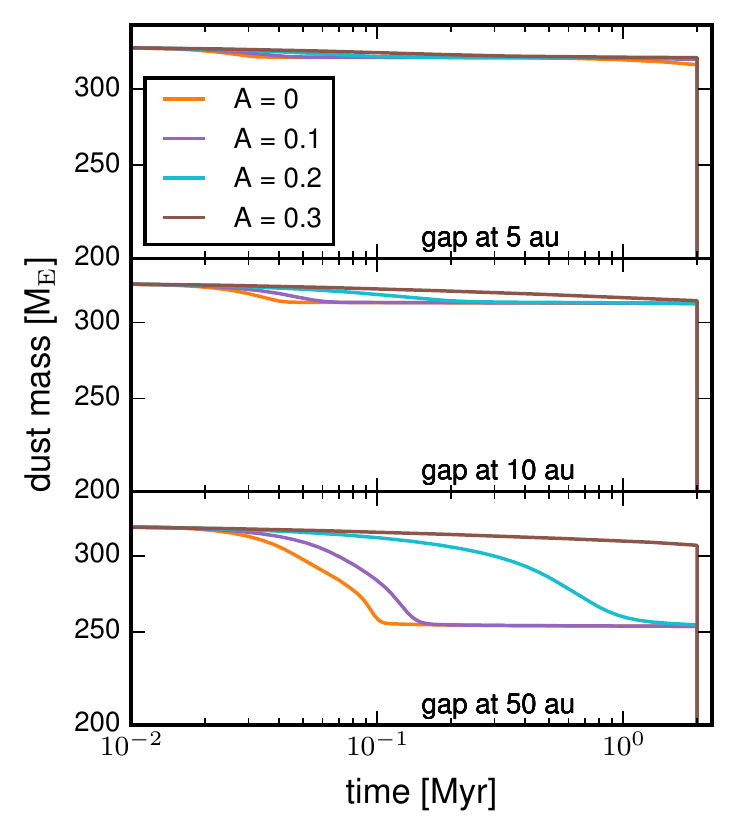}
 \caption{The time evolution of total dust mass in perturbed disc with gaps at 5 au, 10 au and 50 au. With a gap 5 au and 10 au, there is no significant evolution of dust even for the case of A = 0. With a gap at 50 au, total dust mass decays to just about 250 $M_{\rm{E}}$ for A = 0, 0.1, 0.2 but remains fairly unchanged for the case of A = 0.3. for the 2 Myr of disc evolution.}
 \label{fig:fig006}
\end{figure}
\begin{figure*}
 \includegraphics[width=\textwidth]{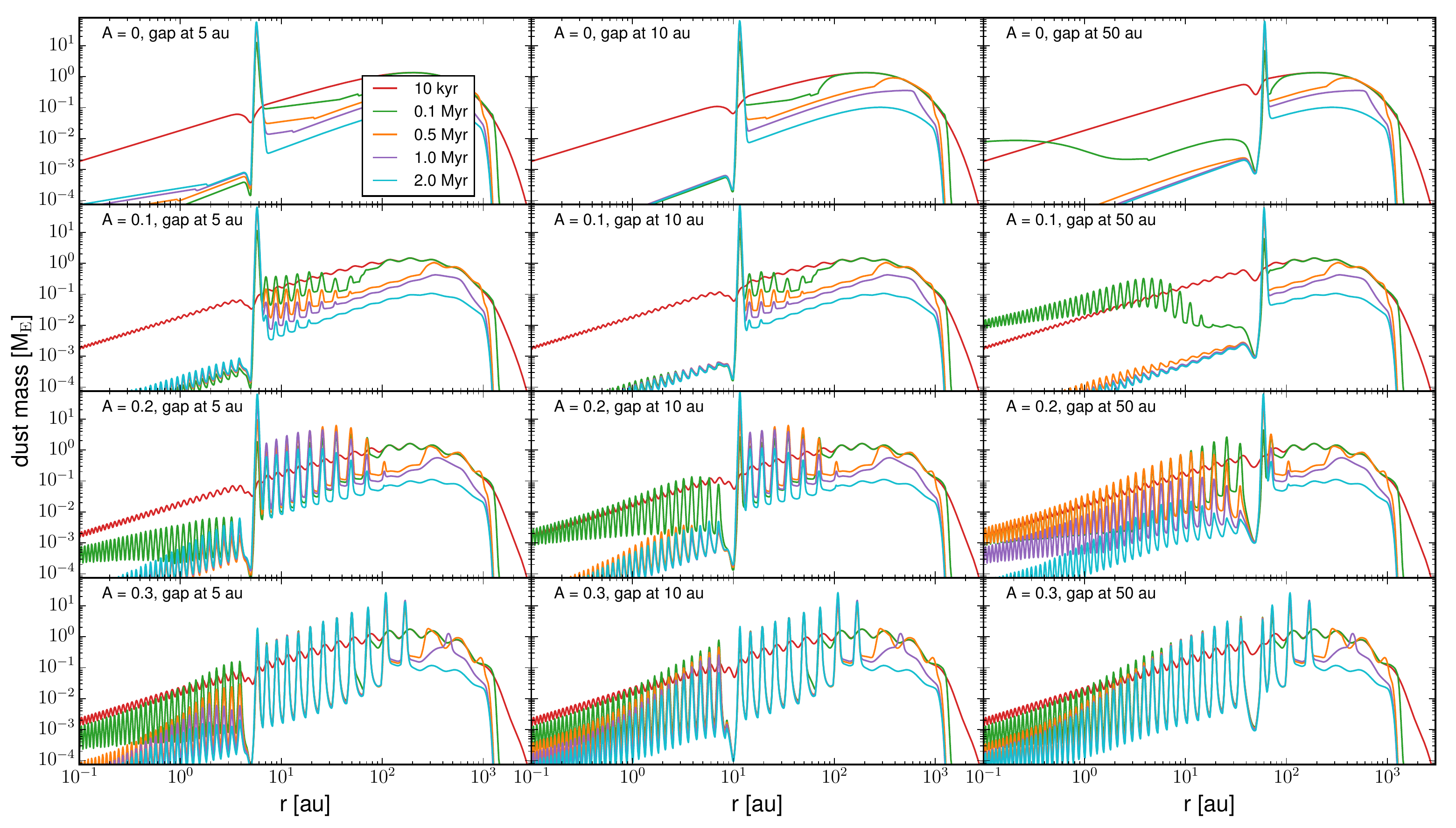}
 \caption{The radial distribution of total dust mass at different evolution times 10 kyr, 0.1 Myr, 0.5, Myrs, 1 Myrs and 2 Myrs for the models shown in Figure~\ref{fig:fig005}.}
 \label{fig:fig007}
\end{figure*}

In Figure~\ref{fig:fig005}, we present the evolution of dust when a single gap deeper than the amplitude of a sinusoidal perturbation is introduced in the disc at radial distances 5 au, 10 au and 50 au, where each row pertains to disc profiles with different perturbation strengths A = 0, 0.1, 0.2, 0.3.

In Figure~\ref{fig:fig005}, the top row shows dust evolution when the gas density is unperturbed. It can immediately be seen that by 0.1 Myr, most of the dust has considerably drained out of the disc region interior to the gaps at 5 au and 10 au. To begin with, by virtue of grain growth that is known to be an efficient process, dust is rapidly converted into mm -- cm sizes and start to experience rapid inward drift on short dynamical timescale~\citep{weidenschilling1980,nakagawa1986,dullemond2005,brauer2008,birnstiel2010}. Next, the region interior to the gap is not well replenished with dust from the outer disc parts because the gaps create much stronger pressure bumps which act as a barrier to the inward drifting dust grains. 

In the case of a gap at 50 au, the grains drift on a much longer timescale and hence take more than 0.1 Myr before depletion because of the longer distance over which they drift compared with the gaps at 5 au and 10 au. However, in all cases in the top row of Figure~\ref{fig:fig005}, traces of dust grains with surface densities below 2 g/cm$^2$ remain in the regions interior to the gaps where marginal grain growth takes place among the trace population during the rest of disc evolution.

The second row of Figure~\ref{fig:fig005} shows simulations performed in composite disc profile that features both a weaker sinusoidal perturbation with A = 0.1 and a deeper gap. Again the grains are lost quickly in the same time frame as the case for A = 0. As already discussed before, this is because a perturbation with A = 0.1 induces weak positive pressure gradients that have little effect on inward migration of grains.

An increase in the wave amplitude to A = 0.2, as shown in the third row of Figure~\ref{fig:fig005}, delays the inward migration of dust grains in the first 0.1 Myrs when compared with A = 0.1 since the stronger bumps initially block the larger grains. However, despite the delayed inward drift of grains caused by the bumps, the surface densities fall very much below that of the unperturbed case within 0.1 Myr, especially for the gaps at 5 au and 10 au. For these two gaps, the delay in inward drift of grains is just temporary where it can be seen that after 0.5 Myr, the grain surface density has strongly dropped. This is because at first the grains get trapped in the pressure bumps after growing to larger sizes within 0.1 Myr. After this, turbulent mixing may force the trapped grains to fragment into smaller sizes and consequently diffuse across the pressure bumps where they migrate inward within 0.5 Myr.

However, for the gap at 50 au,  dust retention is possible between 1 -- 50 au only in the first 0.1 Myrs. One reason for this could be that there are no prominent waves to the right of the pressure bump due to the gap at 50 au  that would hold up grains much like in the cases for the gaps at 5 au and 10 au. Another reason could be that grain trapping at the pressure maxima from the 50 au gap during this period could be inefficient at first as grain growth might be still ongoing. Hence the smaller grains could break through the pressure bump and are transported to the inner parts of the disc.

In the last row of Figure~\ref{fig:fig005} where A = 0.3, the dust grains are able to break through the pressure bumps only in disc regions inside 1 au where the dust surface densities fall below the unperturbed profile. However, roughly outside 1 au, grains are heavily trapped for the rest of the 2 Myr over which the disc evolves. As mentioned before, wave amplitude A = 0.3 is ultimately strong enough to effectively trap most dust grains as early as 0.1 Myr. Hence the presence of a gap in our disc with a perturbation amplitude of A = 0.3 causes much slower grain loss in comparison with weaker amplitudes of A = 0.1 and A = 0.2.

 Figure~\ref{fig:fig006} shows the time evolution of total disc mass corresponding to the simulations in Figure~\ref{fig:fig005}. Here due to the presence of gaps, most of the material is blocked from drifting inward for the case of A = 0, A = 0.1 and A = 0.2, where the amount material that can drift inward depends on the location of the gaps. For example, the amount of material interior to the gaps at 5 au and 10 au is a small fraction of the total disc mass that can eventually flow toward the star and hence the total disc mass does not significantly change after 2 Myrs of disc evolution as shown in the top and middle panels of Figure~\ref{fig:fig006}. For the gap at 50 au, there is a larger fraction of mass interior to the gap that can drift inward. For example, as shown in the bottom panel of Figure~\ref{fig:fig006}, the total disc mass decays from 330 $M_{\rm{E}}$ to about 250 $M_{\rm{E}}$ for the case of A = 0, 0.1 and 0.2, which means only 80 $M_{\rm{E}}$ was able to drift inward after 2 Myrs of disc evolution. However, for A = 0.3, the presence of gaps do not have an effect because  most of the material is already trapped within the strong pressure bumps.
 
As before, to put dust evolution in these different environments in perspective, we present the total dust mass in the disc as a function of radial distance as shown in Figure~\ref{fig:fig007}. In almost all regions to the left of the gaps, dust masses have fallen far below 1 $M_{\rm{E}}$ because dust grains have thoroughly drained out in a very short time as discussed above. Here, only for gaps at 50 au in the simulations with A = 0.2 and A = 0.3 do we obtain total dust masses above 1 $M_{\rm{E}}$, particularly at pressure bumps between 5 au and 50 au. Apparently from Figure~\ref{fig:fig007}, several Earth masses of dust are trapped at the pressure bumps induced by the gaps, and this could be potential sites for planetesimal formation~\citep{drazkowska2016,schoonenberg2017,drazkowska2017,drazkowska2018,stammler2019} and subsequent core growth by planetesimal/pebble accretion~\citep{guilera2017,morbidelli2020,guilera2020,muller2021,izidoro2021}.

In a nutshell, the presence of gaps in the disc can cause considerable grain loss in a short time at a rate that depends on the location of the gap. Gaps that form within 10 au cause faster grain loss via radial drift within 0.1 -- 0.5 Myr even in the presence of pressure bumps that may originate from physical effects such as zonal flows. This has dire consequences for both planetesimal formation and core growth, especially interior to the gap. For example, if a fairly strong pressure bump develops at the water-ice line, typically within 10 au~\citep{lecar2006,min2011,mulders2015,pinilla2016}, it could block the inward drift of pebbles, while at the same time the inner pebbles could be lost over a very short timescale. This could ultimately impede not only core growth via pebble accretion but also planetesimal formation, especially in the inner disc regions $< 5$ au. 

\subsection{Grain size evolution in a perturbed disc without a gap}
\begin{figure*}
 \includegraphics[width=1.125\textwidth]{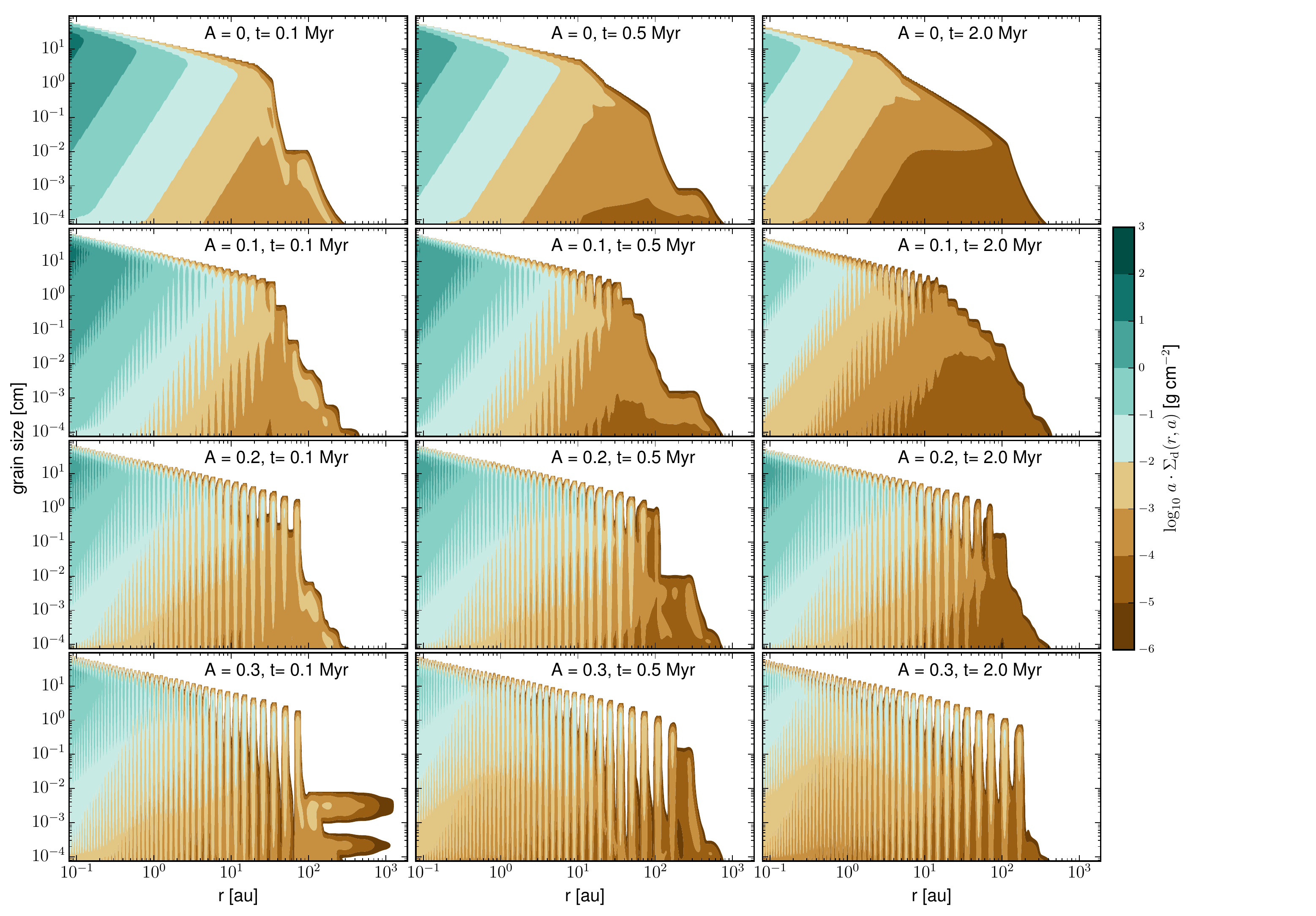}
 \caption{The radial grain size distribution for the models shown in Figure~\ref{fig:fig002}. The left, middle and right columns show the grain size evolution reconstructed at 0.1 Myrs, 0.5 Myrs and 2 Myrs, respectively. The first, second, third and fourth rows show grain size distributions for A = 0 (unperturbed disc), A = 0.1, A = 0.2 and A = 0.3 respectively.}
 \label{fig:fig008}
\end{figure*}
\begin{figure*}
 \includegraphics[width=\textwidth]{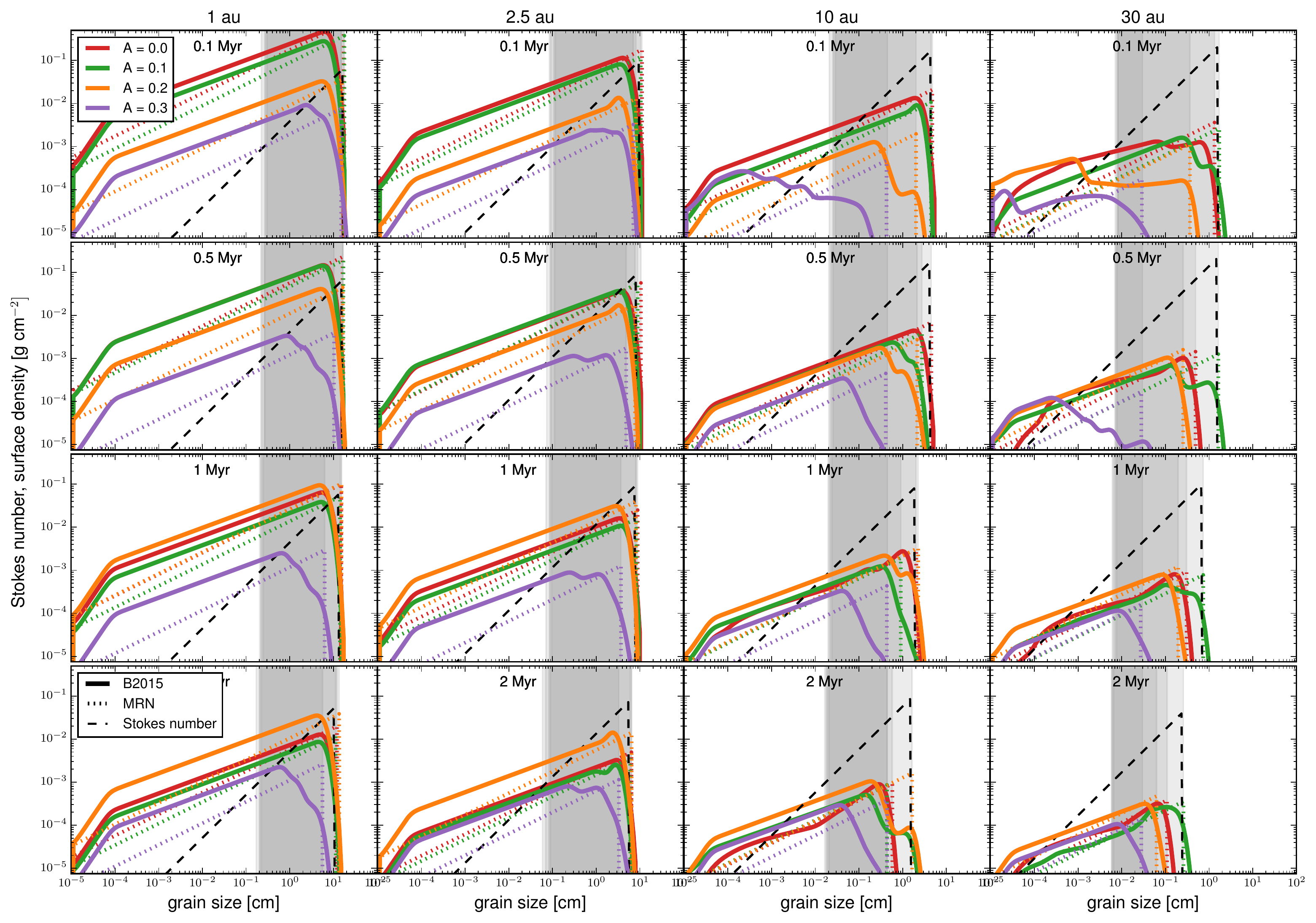}
 \caption{The reconstructed grain size distribution at radial distances 1 au, 2.5 au, 10 au, 30 au, which are the locations of pressure minima, where the grain sizes were reconstructed times 0.1 Myrs, 0.5 Myr, 1 Myrs and 2 Myrs for the dust evolution shown in Figure~\ref{fig:fig008}. The shaded parts indicate the typical range of grain sizes that we used in core accretion where the Stokes numbers range between 0.001 -- 0.1. Here, {\it{B2015}} is the distribution obtained using the recipe in~\citet{birnstiel2015}. The MRN distribution is also shown for purposes of comparison between the two grain distribution models.}
 \label{fig:fig009}
\end{figure*}
\begin{figure*}
 \includegraphics[width=\textwidth]{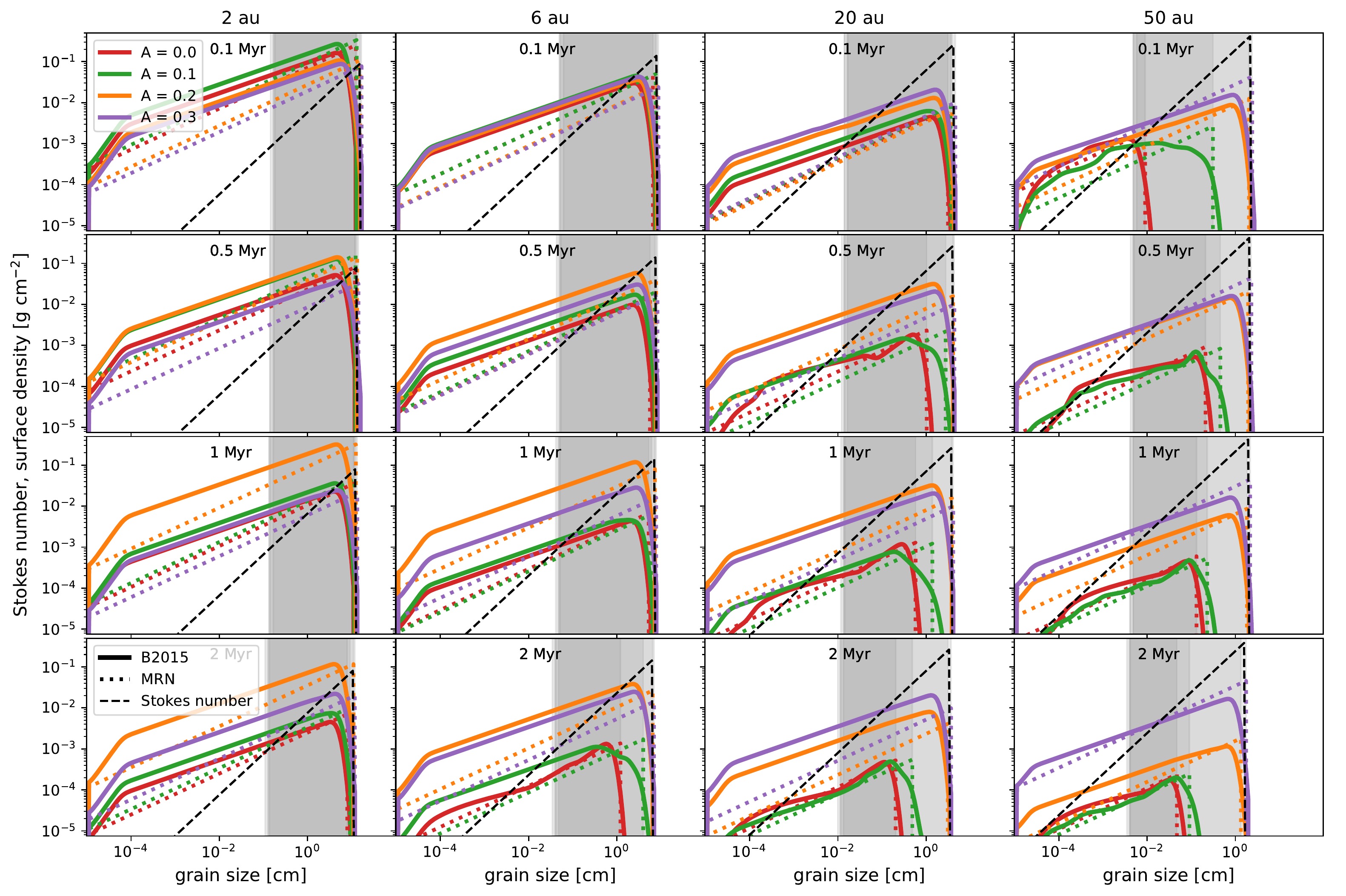}
 \caption{The reconstructed grain size distribution with the same meaning as in Figure~\ref{fig:fig009}, but at radial distances 2 au, 6 au, 20 au and 50 au that lie inside pressure bumps. }
 \label{fig:fig010}
\end{figure*}
The grain size distributions in both unperturbed and perturbed discs in our simulations are shown in Figure~\ref{fig:fig008}. In the top row of Figure~\ref{fig:fig008}, we first present the radial grain size distribution in an unperturbed disc. Here, after 0.1 Myr of disc evolution, a large fraction of dust has already been converted into  mm--cm size grains that have drifted inside 10 au, but submillimetre dust species are still present in large parts of the disc. However after 0.5 Myr, continued grain growth results into considerable depletion of the submillimetre grains in the outer disc regions > 10 au. Further conversion of the smaller grains continues until 2 Myr of disc evolution during which most of the larger grains have drained out from the disc.

For a sinusoidal perturbation with A = 0.1, as shown in the second row of Figure~\ref{fig:fig008}, the grain size evolution differs only slightly from the unperturbed distribution (top row). This is because, with low perturbation amplitude of A = 0.1, even large grains may easily overcome the pressure bumps via turbulent diffusion. However, for the case of larger amplitudes, where A = 0.2 and A = 0.3 (see the third and last rows of Figure~\ref{fig:fig008}), most of the grain species are locked up inside the pressure bumps with marginal inward drifts. After 0.1 Myr, grains might have reached growth/fragmentation equilibrium in most parts of the disc as growth continues slowly. This is because, in addition to fragmentation, grain growth and hence size distribution are also regulated by the available material locked up within the pressure bumps. 

In the last two rows of Figure~\ref{fig:fig008}, there is a minimal change in grain size distribution soon after 0.1 Myr. This is probably because most of the grain species might have reached their growth and fragmentation equilibrium within each pressure bump. Also when these grains have grown to sizes at which they cannot easily diffuse out of the pressure bumps, they are held up there for long periods of time. However, smaller grains that are much more coupled to the gas may still be transported by the gas, and as they grow by collision to larger sizes they may in turn be trapped at the pressure bumps. 

From Figure~\ref{fig:fig008}, we note that for the case of A = 0.3, there is finger-shaped dust distribution in the very outer disc regions that extend beyond 100 au, where there is a gap in the grain sizes roughly between 2 – 20 $\mu$m. The same feature can be seen in Figure~\ref{fig:fig011}.  
This may be explained by the schematic of Figure 1 in~\citet{birnstiel2015}, where particles grow towards either the fragmentation or drift size limit. At wider orbital distances such as 100 au and beyond, fragmentation is not effective so that the grain size distribution is strongly top-heavy in larger grains with much reduced small dust grain population~\citep{birnstiel2015}, leaving a gap between the top-heavy and small size distributions as shown by the finger-shaped distributions in Figure~\ref{fig:fig008}. However, after some time, radial diffusion transports small grains outwards from the inner disc, radially smearing out the the top-heavy distribution~\citep{birnstiel2015},  which makes the finger-shaped distributions disappear after 0.5 Myrs as shown in Figure~\ref{fig:fig008} for the case of A = 0.3. From Figure~\ref{fig:fig008}, the finger-shaped distributions are not manifested in the models with A = 0, 0.1, 0.2 possibly because the radially mixing could have occurred more quickly in a few thousands of years compared with the strongly perturbed disc with A = 0.3, which could have significantly slowed down radial diffusion across the bumps.

In Figures~\ref{fig:fig009} and~\ref{fig:fig010}, we show grain size distribution at selected locations of pressure minima (1 au, 2.5 au, 10 au, 30 au) and pressure maxima (2 au, 6 au, 20 au, 50 au), respectively. In Figure~\ref{fig:fig009}, the grain surface densities with for A = 0.3 at all the locations of pressure minima are far below that A = 0, 0.1, and 0.2. This is because as, discussed before, there is virtually no inward drift of dust in the strongly perturbed disc. However, for A = 0.2, the grain surface densities are close that of A = 0 and A = 0.1, after 1 Myr of disc evolution because after this time dust material begins to start drifting significantly toward the host star. For the cases of A = 0 and A = 0.1, the evolution of grain surface densities are similar, which decay with time, as discussed before.

In Figure~\ref{fig:fig010}, the reverse trend is observed at the pressure maxima, where the grain surface densities for the cases of A = 0.2 and A = 0.3 remain high for extended period of time while the surface densities for A = 0 and A = 0.1 decay as in the previous case in Figure~\ref{fig:fig009}. 

In both Figures~\ref{fig:fig009} and~\ref{fig:fig010}, we overplotted the grain size distribution that would result from the MRN model~\citep{mathis1977}, where the distribution is described by $n(a)\propto a^{-3.5}$. 
Here, the MRN model gives a steeper slope in the grain surface density than in the case of~\citet{birnstiel2015}. This mismatch possibly comes from the fact that MRN model is a single power law model which does not take grain growth, fragmentation and drift into account as in the more complex version of~\citet{birnstiel2015}, which considers different size regimes. Nevertheless, in some cases such as A = 0 and A = 0.1 at 10 AU, the two models are in very close agreement as can be seen in Figure~\ref{fig:fig009}. The influence of grain growth, fragmentation and drift on the distributions can be seen from Figures~\ref{fig:fig009}, where the two models take different shapes at 30 au especially in the case of A = 0.2 and A = 0.3. Here, similar distribution patterns between the two models are obtained depending on how fast grain growth/fragmentation equilibrium is attained. For example, from Figures~\ref{fig:fig009}, for the case of A = 0.2 and A = 0.3 at 30 au, the two distributions take similar shapes after 0.5 Myr and 1 Myrs, respectively. This is because grain growth might have proceeded slowly before reaching growth/fragmentation equilibrium. For the case of pressure maxima shown in Figure~\ref{fig:fig010}, the distributions from both models follow the same patterns for the respective perturbation strengths for all the evolution times considered. This is possibly because most grains have drifted to the pressure maxima where growth/fragmentation equilibrium is achieved quickly.
\subsection{Grain size distribution in a perturbed disc with a gap}
\begin{figure*}
 \includegraphics[width=1.125\textwidth]{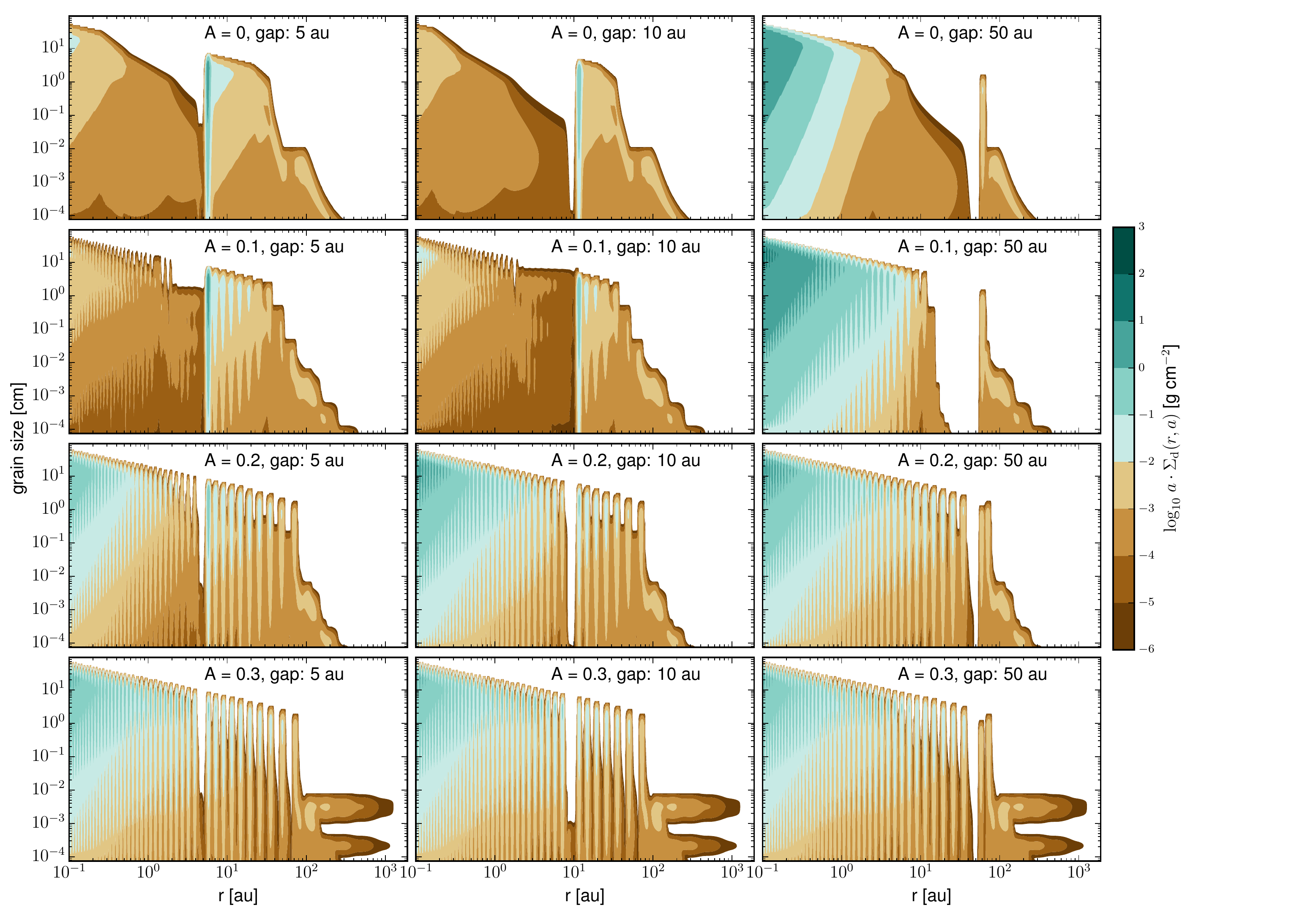}
 \caption{The radial grain size distribution for the disc models which include both sinusoidal pertubation and a Gaussian gap profile shown in Figure~\ref{fig:fig005} where the grain size evolution is reconstructed at 0.1 Myrs. The first, second, third and fourth rows show grain size distributions for A = 0 (unperturbed disc), A = 0.1, A = 0.2 and A = 0.3 respectively.}
 \label{fig:fig011}
\end{figure*}
Figure~\ref{fig:fig011} shows trace species of dust grains that have evolved for 0.1 Myr in our disc model featuring both a perturbed gas density and a gap at radial locations 5 au, 10 au and 50 au. In our simulations in Figure~\ref{fig:fig011}, although sinusoidal perturbations in the density structure may help retain grains, a simple introduction of a gap deeper than the amplitude of the sinusoid results into substantial loss of grains within just 0.1 Myr. This also demonstrates the fact that grain growth is generally a very rapid process which, if not regulated by some other process such as fragmentation, can lead to significant loss of grains via radial drift.

For the unperturbed density profile in the top row of Figure~\ref{fig:fig011}, grains of almost all sizes interior to the gap, and most notably those with sizes greater than 1 cm, have been lost through radial drift in just less than 0.1 Myr. However, for the density profile with a gap introduced at 50 au, mm -- cm particles are still present at 0.5 Myr within 1 au compared with the profiles with gaps at 5 au and 10 au. This is partly because, on one hand, particles that grow at long distances as far as 50 au take much longer time to drift to the inner disc region. On the other hand, particles that grow to fragmentation and drift limits within 10 au cover shorter distance and time before they are lost through radial drift. 

As seen in the top panel of Figure~\ref{fig:fig011}, dust accumulates at the high peak pressure bumps to the right of the gaps at 5 au and 10 au than at 50 au. This is because, at first, smaller grains that are more coupled to the gas are carried along with the gas across the high peak pressure bump that is to the right of the gap at 50 au. Then after crossing the pressure bump, these smaller grains may grow large and drift to the inner parts of the disc. Also, only a few grains that are able to grow to larger sizes beyond 50 au get blocked at the pressure bump. Furthermore, more dust grains grow to larger sizes in large parts of the disc beyond the 5 au and 10 au gaps from where they drift and get blocked at pressure bumps due to these gaps.

In the second row in Figure~\ref{fig:fig011}, dust size distribution is occasioned by a sinusoidal perturbation with wave amplitude of A = 0.1, which results a significant loss of grains similar to the unperturbed case. However, in this setup, more of the larger grains are trapped at the pressure bumps compared with the unperturbed density profile. For example, most grains that have grown to over 1 cm are now partially locked within the sinusoidal pressure bumps as compared with the case for A = 0. While traces of mm-cm size grains remain interior to the gaps at 5 au and 10 au after 0.1 Myr of disc evolution, the evolution of dust in a perturbed disc with a gap at 50 au features mm -- cm grains that dominate grain size distribution at the pressure bumps within 5 au.

In the case of stronger perturbation with A = 0.2 and A = 0.3 as shown in the third and fourth rows of Figure~\ref{fig:fig011}, grain sizes are much larger with higher surface densities than the waves with amplitudes A = 0 and A = 0.1. This is because with a stronger amplitude, particles undergo weaker turbulent diffusion across strong pressure bumps. Another reason could be that since the smaller grains follow the gas streamlines, they drift less slowly since the gas in our model has not evolved much within 0.1 Myr. These smaller grains can then grow to larger sizes, thereby increasing the amount larger grains within the pressure bumps. From the bottom row of Figure~\ref{fig:fig011}, there are still some traces of submillimetre to millimetre sized dust species beyond 100 au, which shows that these smaller grains grow much more slowly within 0.1 Myr. This may be attributed to the low collision velocities and low surface densities at these wider distances.

From Figure~\ref{fig:fig011}, for the case of A = 0.1, the distribution shows a wider gap at 5 au and 50 au. This is possibly because for the weakly perturbed disc where A = 0.1 with a gap at 50 au, the inward drift of dust grain is delayed by the pressure bumps beyond 50 au, which results in poor replenishment of grains inside 50 au, causing a bigger deficit of grains interior to 50 au.  Here, the grains interior to 50 au drift faster than they can be promptly replenished, hence resulting in a wider gap. This is not the same case with A = 0, A = 0.2 and A = 0.3. First of all for the case of A = 0, there are no pressure bumps that can cause a delay in inward drift of grains from disc regions outside 50 au and hence there is almost immediate replenishment of material interior to 50 au from dust grains that are able to penetrate the pressure bump due to a gap at 50 au. 
This can easily be seen from the profiles of dust masses with gaps at 50 au shown in Figure~\ref{fig:fig007}, where between 20 -- 50 au, at 0.1 Myr, the dust mass in this region for the case A = 0.1 is smaller than that of A = 0, which means that this region was not replenished at the same rate as for A = 0. A similar trend can be observed for the gaps at 5 au and 10 au,  where there is small dust mass in the interior neighbourhood of 5 au and 10 au for the case of A = 0.1 compared with A = 0, but the gap is not prominent at 10 au. On the other hand, for the cases of A =0.2 and  A = 0.3, the presence of strong pressure bumps in the vicinity of the gap traps dust material where the dust grains do not diffuse easily across the pressure bumps.

\subsection{Core growth in perturbed disc}
\begin{figure*}
 \includegraphics[width=\textwidth]{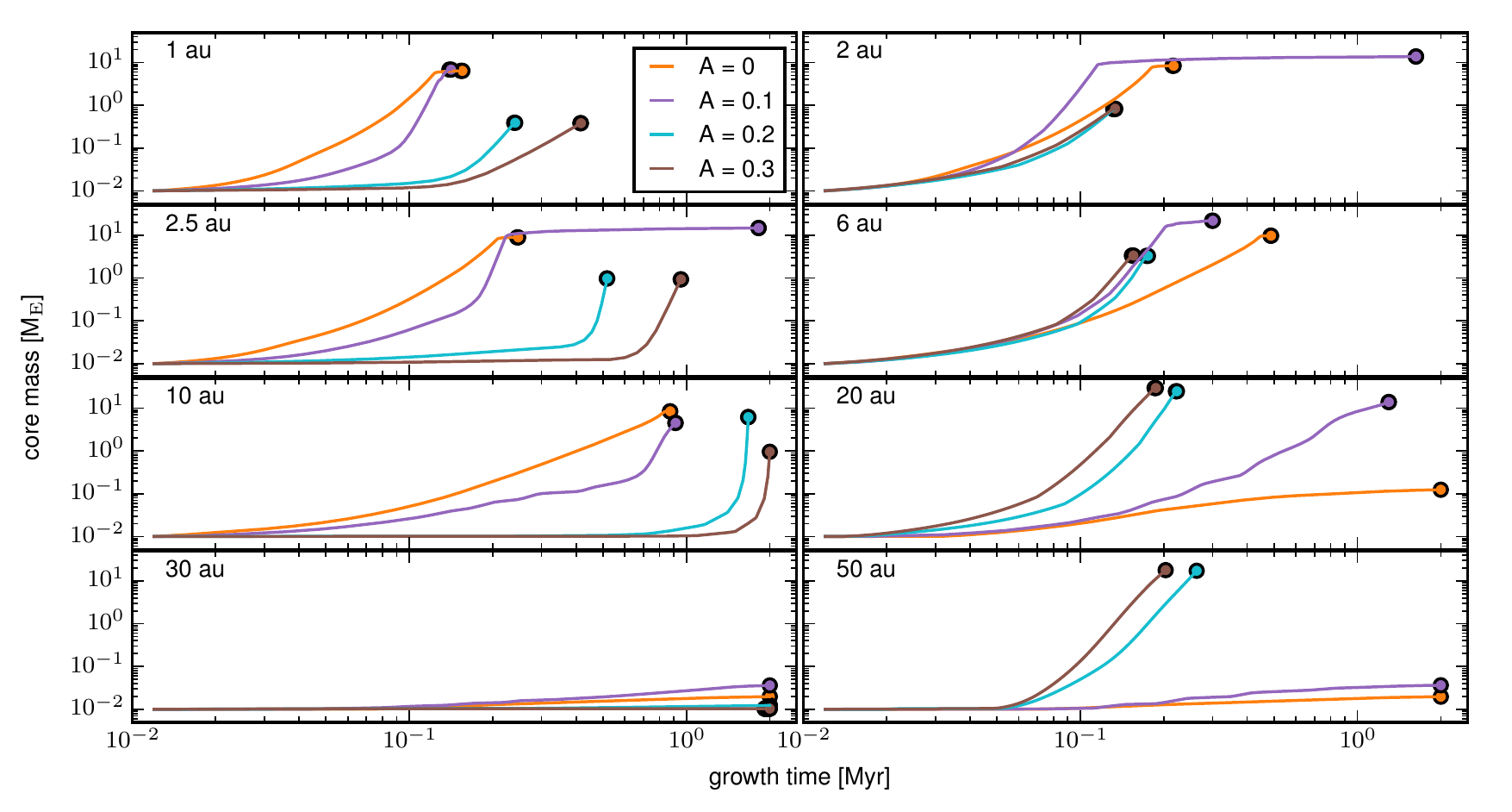}
 \caption{The time evolution of planetary cores in unperturbed and perturbed discs with $u_{\rm{f}}=10~\rm{m/s}$ and $\alpha_{\rm{t}}=10^{-3}$. The cores started accreting pebbles at the different initial positions indicated on each plot. The left and right panels show planetary cores that started accreting at the pressure minima and maxima, respectively.}
 \label{fig:fig012}
\end{figure*}
\begin{figure*}
 \includegraphics[width=\textwidth]{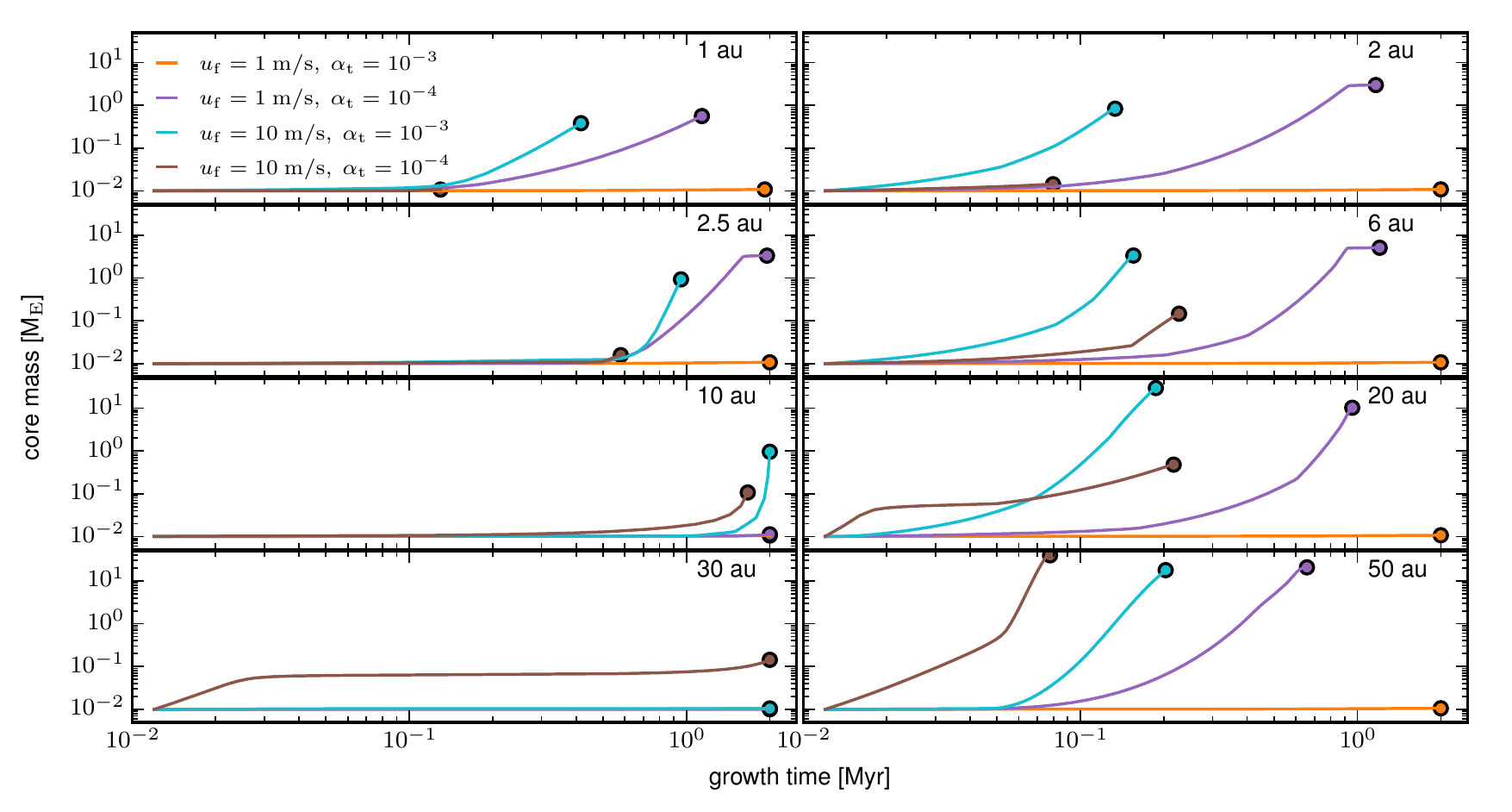}
 \caption{The time evolution of planetary cores in strongly perturbed disc with A = 0.3 for different combinations of fragmentation velocity and turbulence viscosity. Here, the left and right panels show planetary cores that accrete at the pressure minima and maxima, respectively.}
 \label{fig:fig013}
\end{figure*}

\begin{figure*}
 \includegraphics[width=\textwidth]{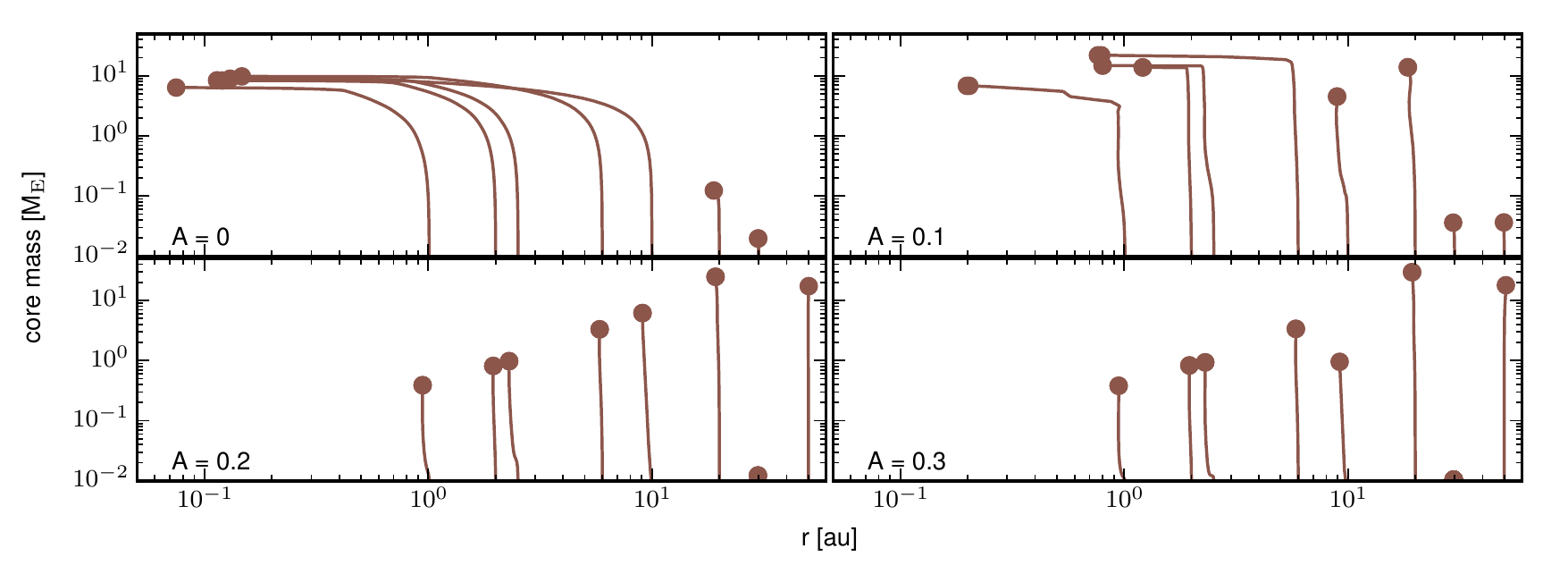}
 \caption{The orbital evolution of the planetary cores shown in Figure~\ref{fig:fig012}.}
 \label{fig:fig014}
\end{figure*}
\begin{figure*}
 \includegraphics[width=\textwidth]{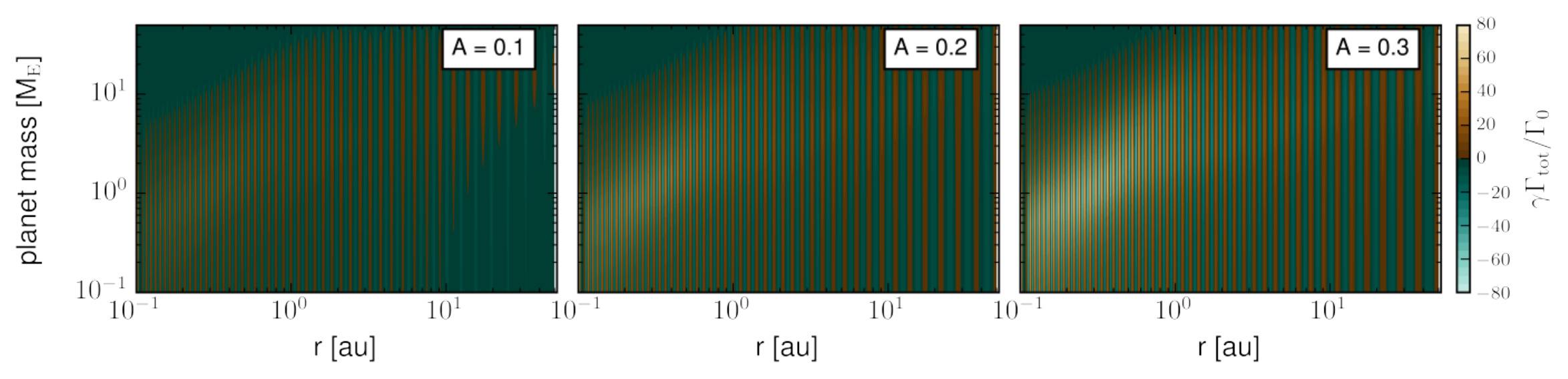}
 \caption{The torque maps in perturbed disc with perturbation amplitudes of A = 0.1, A = 0.2 and A = 0.3. Here, inside and outside the pressure bumps, the net torque on the planet is positive and negative, respectively.}
 \label{fig:fig015}
\end{figure*}

 Figure~\ref{fig:fig012} shows the growth tracks of the planetary embryos that assembled by pebble accretion in both unperturbed (A = 0) and perturbed discs with wave amplitudes A = 0.1, A = 0.2 and A = 0.3. The left panel shows the planetary embryos that were initially deployed at orbital positions 1 au, 2.5 au, 10 au and 30 au, which lie roughly in the pressure minima. The right panel shows planetary cores that started growing inside pressure bumps located at orbital distances 2 au, 6 au, 20 au and 50 au. We fixed the final core mass by the pebble isolation mass if accretion took place outside the pressure maximum. In the case of accretion inside the pressure bump, we stopped core growth upon depletion of the available pebbles trapped inside the pressure bump. In the following discussions, we compare how different perturbation strengths affect time evolution of planetary cores at the different orbital positions. 
 
 In our growth model for the cases of A = 0 and A = 0.1, the planetary core accretes until it reaches the pebble isolation mass of pebble species with the smallest Stokes number in the grain size distribution as in \cite{andama2021}. This is because, as discussed in Section~\ref{dust_waves}, the dust evolution for both cases is similar. Consequently, with weak pertubation of A = 0.1, there is transport of material between the pressure bumps. This material continues to be accreted by the growing planet, until the planet induces its own pressure bump that eventually blocks the inward drifting pebbles. For the perturbed discs with A = 0.2 and A = 0.3, our core masses were constrained by the amount of material available within the pressure bumps because  most of the material is trapped in the pressure bumps unlike for the cases with A = 0 and A = 0.1.

From the left panel of Figure~\ref{fig:fig012},  planetary cores that were placed at the pressure minima initially accreted slowly compared with the growth rates in the unperturbed disc, for which A = 0, where the growth rates drop further with increasing wave amplitudes. This is because the flow of material through the disc is delayed by the pressure bumps that trap most of the dust grains, and this delay increases with the strength of the pressure bumps. Hence, there is less available material between the pressure bumps, which scales down the accretion rate between the pressure bumps according to the strength of the induced pressure maxima. However, with time, the planet migrates inside the pressure bump and starts growing more efficiently, where its final mass is limited by the available material inside the bumps.

However, for cores  that started at 1 au and 2.5 au, the growth times were much shorter compared with those that started at 10 au and 30 au. This can be associated with grain size distributions and their abundances at the respective locations (see Figure~\ref{fig:fig009}). At 10 au and 30 au, the grain distribution is dominated by submillimetre to millimetre sizes with low surface densities, making core growth very slow. At 1 au and 2.5 au, the grain sizes are larger and have higher surface densities, and hence core growth rates are much more efficient, compared with the growths at 10 au and 30 au.

For the case of 30 au, all the planets accreted very small amount of pebbles and remained below 0.1 $M_{\rm{E}}$. One reason for this is that at 30 au only small amounts of mm -- cm size grains are available as shown in Figure~\ref{fig:fig009} for all the disc profiles. Another reason is that this orbital distance is just outside the pressure bump, from where much of the material is attracted away toward the pressure maxima (see Figure~\ref{fig:fig002}). Here, the planets could not migrate inside the bumps where they would find and accrete the trapped material, as was the case for the cores that started at the other locations.

In the right panel of Figure~\ref{fig:fig012}, planetary cores started growing inside the pressure bumps located at the orbital positions 2 au, 6 au, 20 au and 50 au. Here, the core growth times are shortest in the perturbed discs with A = 0.2 and A = 0.3, where the cores took less than 0.2 Myr to reach their final masses, as compared with the growth times of the cores that started at the pressure minima. This is because there is enough material inside the bumps, which facilitates fast core accretion~\citep[e.g.,][]{morbidelli2020}. On the other hand,  the growth times for A = 0 and A = 0.1 are longer. The reason for this is that there is continuous flow of material through the planet's orbit, and the final core masses are determined by pebble isolation mass rather than the amount of material trapped in the bumps, as was the case for A = 0.2 and A = 0.3. Furthermore, not all the material that crosses the planet's orbit is accreted very efficiently, which increases the growth time. In the case of a planet inside a pressure bump, it is likely to accrete most of the material trapped within the bump more efficiently, hence the improved growth times.

Comparing the final core masses, we see that most of the cores  for the case of A = 0.1 are more massive than the case of A = 0, especially for the starting positions $\le$ 20 au.  For example, at 20 au, the core masses in A = 0.1 and A = 0 are 10 $M_{E}$ and 0.1 $M_{E}$, respectively. As another example, at initial orbital position of 6 au, the final core masses are 20 $M_{E}$ and 10 $M_{E}$ for A = 0.1 and A = 0, respectively. One would expect the core growth patterns in both cases to closely follow each other since the total dust mass evolution is presumably similar in both cases, as we saw in section~\ref{fig:fig003}, which is further illustrated by the similar grain size evolution shown in Figure~\ref{fig:fig009}. However, the growth patterns are not necessarily the same because even the weaker pressure bumps due to A = 0.1 may hold more material at a particular location than the unperturbed disc as shown in the right panel of Figure~\ref{fig:fig002}. This may then permit a core to grow bigger than the case in the unperturbed disc.

For the cases A = 0.2 and A = 0.3, all the core masses are pretty the same except at 10 au. For these two cases with the cores placed at 10 au, the planets accreted material very slowly in the first 1 Myrs and reached approximately 20 $M_{\rm{E}}$ and 1 $M_{\rm{E}}$  after 2 Myrs for A = 0.2 and A = 0.3, respectively.  For the case of A = 0.2, the initial slow  growth also matches the late movement of material in the disc which started at $\sim$ 0.8 Myrs as shown in Figure~\ref{fig:fig003}, which then replenishes the pressure bumps. However, for the case of A = 0.3, lack of substantial movement of material between the pressure bumps delays the core accretion until $\sim$ 1.5 Myrs, where the core accreted only 1 $M_{\rm{E}}$ of material. 

Figure~\ref{fig:fig013}  further illustrates the diversity of planetary cores produced in a perturbed disc for the case of A = 0.3. For $u_{\rm{f}} = 1~\rm{m/s},~\alpha_{\rm{t}} = 10^{-3}$, all the planetary embryos failed to grow because for this pair of parameters, most of the grain sizes are very small which are difficult to accrete since they are tightly coupled to the gas~\citep[for detailed discussion see][]{andama2021}.
As we saw before, most of material is trapped at wider orbits where the large grains hardly diffuse to the inner disc regions. This results in smaller and larger cores in the inner and outer disc regions, respectively. We remind the reader that in our model, the outcome of the planetary core depends pretty much on the amount of material trapped at pressure maxima and minima and how much material is able to diffuse across the pressure bumps.
For example, for the case of higher fragmentation velocity threshold of $u_{\rm{f}} = 10~\rm{m/s}$ and low turbulence viscosity of $\alpha_{\rm{t}} = 10^{-4}$, the grain sizes are bigger which do not easily diffuse through the strong pressure bumps, leaving a deficit either at the pressure minima or at locations in the inner disc regions. Hence, at 1 au, 2 au and 2.5 au planetary cores failed to grow because of small availability of large grains at these locations, where planetary cores exhaust these small amounts of pebbles in a short time.

From our results in Figures~\ref{fig:fig012} and~\ref{fig:fig013}, our model does not necessarily favour any choice of parameters, a part from the case where $u_{\rm{f}} = 1~\rm{ m/s}$ and $\alpha_{\rm{t}}=10^{-3}$ where the cores do not grow. In particular, the cores can form early and bigger for any combinations of $u_{\rm{f}}$ and $\alpha_{\rm{t}}$ chosen in this study, especially when the growth takes place at the pressure bumps. The size of planetary cores and growth times in our model depends strongly on how much material is trapped within the bumps at a particular location in the disc. In general core growth is rapid at the pressure bumps which takes as short time as 0.2 Myrs, where the final mass depends on the amount of material trapped in the pressure bumps, explained in detail in Section~\ref{implications}.

In Figure~\ref{fig:fig014}, we present the orbital evolution of planetary cores of Figure~\ref{fig:fig012}.
In the unperturbed disc where A = 0, planets that start growing inside 10 au migrated significantly. The same trend is seen in the perturbed disc with A = 0.1, but in this case the planets migrate over a lesser distance compared with A = 0. For example, a planet that started accreting at 6 au in the unperturbed disc migrated to about 0.1 au but in the case of A = 0.1, the planet migrated to about 1 au. For A = 0.1, the planets initially grow nearly in-situ inside the pressure bump since the cores are small and exert less torque on the gas disc. Also inside the pressure bumps, the net torques are positive as shown in Figure~\ref{fig:fig015}, which prevents inward migration of the planets. However, the planets later migrate inward from the pressure bumps as they grow more massive, which implies the corotation torque could have saturated which then forces the wave torques to push the planet inward. However, from 10 au and beyond, the planetary cores are locked up inside the pressure bumps and hence do not migrate. 

As shown in Figure~\ref{fig:fig014}, with increased strength of pressure bumps, the cores hardly migrate and are trapped indefinitely in the pressure bumps because of the nearly zero net torque on the planet, as illustrated in the torque maps of Figure~\ref{fig:fig015}. However, the planet may open a gap and alter the gas surface density as well as the initial density profile inside the pressure bumps. Therefore, the effects of wave and corotation torques may then change, causing the planet to migrate away from the pressure bump. The exact nature of planet migration in already perturbed discs and the impact of gap opening by the planet on the gas surface density were not within the scope this study and may require further investigation in future studies. Nevertheless, Figure~\ref{fig:fig015} suggests that changes in the pressure gradient may greatly reduce or reverse the migration of the planet, potentially saving the cores from being lost to the central star.

\subsection{Implications for planet formation}\label{implications}
Clearly the unperturbed and perturbed discs play different roles in the planetary core growth patterns in our simulations. First of all, in unperturbed discs there is constant flow of solid material through the orbit of the planet, which sustains core growth via pebble accretion. At the same time, the planet may migrate to regions where material is abundant, which further facilitates their growth. 

Secondly, in the perturbed discs, the pressure bumps might prevent planetary cores from migrating to other bumps where more material can be found and hence the planets only accrete inside the pressure bumps within which they are trapped (see Figure~\ref{fig:fig012}). Therefore, the pressure bumps promote nearly in-situ formation of planets, which can be beneficial for forming cores of giant planets if huge amounts of pebbles are trapped inside the bumps. This is also beneficial in limiting fast type-I migration, which may cause loss of cores before they start accreting gas to become gas giants.

Thirdly, the final core mass of a planet growing inside a pressure bump is dictated by the amount of material that is held up in the pressure bumps. This is because the planet cannot accrete more material than what is trapped in the bumps~\citep{morbidelli2020}. That is, the core masses are limited by the amount of material trapped in the bumps instead of the classical pebble isolation mass, in which case core growth stops as soon as the material insided the pressure bump is exhausted. In our simulations, this holds for strong perturbations in the gas density structure, where most of material remains confined inside the pressure bumps. In fact, we can use the right panel of Figure~\ref{fig:fig002} to check that the corresponding masses temporally confined at 1 au, 2 au, 2.5 au, 6 au, 10 au, 20 au, 30 au and 50 au give a good approximation to the planet masses for the case of A = 0.2 and A = 0.3. The planet may  then consume the trapped material quickly in a short time as shown in Figure~\ref{fig:fig012}, where core growths in the perturbed discs are completed in a shorter time than the respective growth times in the unperturbed disc.

Fourthly, the amount of material and its life time in the pressure bumps depends on the strength of the perturbation among other factors. For instance, if the pressure bump is small, material can easily leak out of the bumps and hence disc evolution may be similar to the unperturbed case. However, the cores may grow more massive in weakly perturbed disc as is the case for A = 0.1. On the other hand, stronger perturbations may utterly prevent inward drift of pebbles, and depending on the amount of material that is initially trapped, planetary cores may grow small or big as shown in the bottom panel of Figure~\ref{fig:fig014}. 

Lastly, a simple introduction of a gap in the disc can cause a rapid depletion of the solids interior to the pressure bump due to the presence of the gap. For example, if a giant planet forms as early as 0.1 Myr, the pressure bump it induces will block inward drift of dust grains. The dust grains in the regions interior to the planet’s orbit are quickly lost to the central star on short dynamical timescales as we demonstrated in Section~\ref{sec:gaps}. This may impede the formation of planets by core accretion of pebbles in these inner disc regions. This may also impede the formation of planetesimals in the interior disc regions since their formation requires dense concentrations of pebbles. We can thus conclude that pressure bumps may cut-off formation of planets completely under some particular conditions. 
Even in the absence of a giant planet, any phenomenon that induces a strong pressure bump can ultimately affect growth of planetary cores interior to the pressure bump. This could be due to viscosity transitions at water-ice line that could lead to pressure bumps. However, it is unlikely that water-ice lines can create strong perturbations in the gas surface density profile that can result in strong pressure bumps, unless viscosity transitions are extreme~\citep{bitsch2014b}.

\subsection{Limitations}
We outline some of the key limitations of our model. To begin with, our simulations lack a model of dust drift that might cause feedback loop on disc viscosity and hence trigger viscous instability in the disc~\citep{hasegawa2015, dullemond2018a,delage2022}. This would require a more sophisticated disc viscosity model as in~\cite{dullemond2018a}, which is beyond the scope of this work. Viscosity transitions could originate from small amounts of dust that may reduce gas conductivity, which suppresses MRI and reduces turbulence in some disc regions where surface density increases resulting in local pressure maxima~\cite{dullemond2018a}. 

Our model is rather an oversimplification of multiple rings and gaps which may not be realistically sinusoidal in nature. 
Nevertheless, under some appropriate physical conditions, the gas disc may self-organise into zonal flows, creating density rings and hence pressure bumps, whose distribution may be wavy~\citep{kunz2013,bai2015, bethune2016,bethune2017, riols2019}, which is the main motivation of our work. 

We have not tested the possibility of planetesimal formation at the bumps which could remove grains from the disc, especially within the bumps, which means reduction in the amount of grains that could fragment and drift again through pressure bumps. Thus, if much of the grains are converted into planetesimals in the bumps, then pebble accretion could  further be impacted since fewer pebbles would be available for accretion. 

The gas surface density in our simulations is static over the 2 Myrs of disc evolution. However, average disc lifetimes constrained by observations are typically in the range of 2 -- 5 Myr, but the actual disc lifetimes could range from 1 --10 Myr~\citep{hartmann1998, hartmann2016, haisch2001, mamajek2009}. Moreover most stars are born in clusters, which may lead to early dissipation of disc due to external photoevaporation.

We assumed a simple temperature profile, and hence the picture of our model may change in radiative discs. For example, the small grain population may significantly contribute to opacity and hence the temperature structure, which in turn affects the gas structure as examined in~\cite{savvidou2020}.

We used a uniform fragmentation velocity throughout the disc, which may not necessarily be the same in all parts of the disc. For example, beyond the water-ice line, grains are predominantly icy and fragment at higher velocities of 10 -- 80 m/s~\citep{blum2008,wada2013,gundlach2011,gundlach2015}. On the other hand, inside the ice line the icy grains sublimate into silicate grains, which fragment at lower velocities of 1 -- 10 m/s~\citep{wada2013,gundlach2015}. At the same time, an opacity transition is generated at the ice line, which may then induce a pressure bump. 

Another important limitation of our model is that we did not model the mechanism of type-I migration in perturbed discs, but rather used the simplified formalism in \cite{paardekooper2011}. Gravitational interaction of the planet with the disc might change the disc structure where the planet might launch its own density waves that could blend with waves originating from other sources. Consequently, this adds extra complication to the disc evolution, which can change the nature of the originally perturbed disc. Thus, detailed hydrodynamic simulations are needed to study how a growing planet changes perturbed discs and to better understand the migration paths in such environments.

Overall, despite the above mentioned limitations, our simplified approach provides a hint to the role played by multiple dust rings and wave-like density perturbations on planet formation, especially by core accretion of pebbles. Future studies should consider addressing the above mentioned limitations of our model to a give a more complete and consistent view of planet formation in environments in which the gas surface density deviates from the smooth profile commonly assumed in many simulations.

\section{Conclusion}\label{conclusions}
We have studied dust evolution in globally perturbed discs where we employed the perturbation schemes described in~\cite{pinilla2012} and \cite{dullemond2018b}. We then carried out numerical simulations to explore how core growth by pebble accretion is affected in the perturbed discs. Our dust evolution routines are based on the two-population code of~\cite{birnstiel2012} that features coagulation, fragmentation and drift limits, where we reconstructed the grain sizes using the reconstruction tool of~\cite{birnstiel2015}.

Grain retention within the pressure bumps depends sensitively on perturbation amplitude, in agreement with~\cite{pinilla2012}. For weak perturbation levels (in our case A = 0.1), the evolution of total dust mass with time closely follows that of the unperturbed discs (A = 0) because the weak pressure bumps do not necessarily stop grain migration. However, in the extreme case of strong perturbation, for example, with A = 0.3, there is virtually no radial movement of solid material in the disc since the grains cannot easily overcome the pressure bumps.

If we introduce a gap in a smooth disc anywhere between 1 -- 50 au, dust interior to the gap locations is rapidly lost within 0.1 Myr, consistent with previous studies~\citep[e.g.,][]{whipple1972,weidenschilling1977,takeuchi2005, alexander2007,brauer2007, brauer2008, johansen2019}.

In the presence of both a gap and a sinusoidal perturbation, where the pressure bump due the gap is stronger, rapid grain loss also occurs on timescales between 0.1 -- 0.5 Myrs, depending on the strength of the wave amplitudes. For example, for a weak perturbation, grains are lost on the same timescale as in the unperturbed disc. Strong wave amplitudes may only delay grain loss if the gap is introduced at wider orbits $\sim 50$ au, because the grains that are initially trapped in the bumps take time to migrate inward. In this case, core growth might still be possible inside the pressure bumps.

The presence of a strong pressure bump in the disc could therefore be a serious problem not only for formation of planetesimals, but also for the formation of planetary cores through the core accretion paradigm, especially in the inner disc regions. In the first place, planetesimal formation requires high particle concentrations, for instance via streaming instability~\citep{johansen2007}, which may not be achievable if grain migration timescales are shorter than the time taken by the grains to reach overdensities for manifestation of gravitational collapse. This means that planetesimal formation must be quick before the grains drift away from the planet(esimal) forming region. Secondly, if planetesimal formation can beat the grain loss timescale, then core growth via pebble accretion must also proceed very fast. Otherwise, an embryo would require another growth path such as accretion of planetesimals, which of course must have formed early enough and plenty in number.

Multiple pressure bumps may restrict core formation by pebble or planetesimal accretion at specific orbital distances where the bumps are located. This is because most of the solid material is attracted toward the pressure bumps, while the regions between the pressure bumps could be heavily depleted, depending on the scale of the pressure gradients. Hence, they could play an important role in the orbital architecture of planetary systems as well as their core masses. Firstly, the core masses are limited by amount of material trapped in the bumps instead of the classical pebble isolation~\citep{morbidelli2020}. Secondly, the bumps may reduce or entirely prevent inward migration of the planets, which could solve the dilemma of rapid inward migration of gas giants in addition to some of the solutions provided in previous studies~\citep[e.g.,][]{paardekooper2014,crida2017,crida2017b,kanagawa2018,robert2018, bergezcasalou2020, ndugu2021}. Thirdly, the bumps can promote core accretion at locations where accretion would be inefficient in a smooth disc, particularly in the outer disc regions.

\section*{Acknowledgments}
We thank the anonymous referee for the very useful comments that helped in improving the paper. We thank Swedish International Development Cooperation Agency (SIDA) for financial support through International Science Programme (ISP) Uppsala University Sweden  to the East Africa Astronomical Research Network (EAARN).
\section*{Data availability}
For the purpose of reproducibility, the code used to obtain results in this paper will be provided upon reasonable request.


\bibliographystyle{mnras}
\bibliography{Andama_et_al_2022.bib} 





\bsp	
\label{lastpage}
\end{document}